\documentclass{article}
\usepackage{amssymb}

\usepackage{graphicx}
\usepackage{amsmath}

\input{tcilatex}

\begin{document}

\author{V. S. Borisov\thanks{%
E-mail: vyacheslav.borisov (at) gmail.com} \\
\\
Department of Mechanical Engineering, Ben-Gurion University, \\
Pearlstone Center for Aeronautical Engineering Studies, \\
Beer-Sheva, Israel}
\title{On dynamics of geometrically thin accretion disks }
\maketitle

\begin{abstract}
Axisymmetric accretion disks in vicinity of a central compact body are
studied. For the simple models such as vertically isothermal disks as well
as adiabatic ones the exact solutions to the steady-state MHD
(magnetohydrodynamic) system were found under the assumption that the radial
components of velocity and magnetic field are negligible. On the basis of
the exact solution one may conclude that vertically isothermal disks will be
totally isothermal. The exact solution for the case of adiabatic disk
corroborates the view that thin disk accretion must be highly nonadiabatic.
An intermediate approach, that is between the above-listed two, for the
modeling of thin accretion disks is developed. In the case of non-magnetic
disk, this approach enables to prove, with ease, that all solutions for the
midplane circular velocity are unstable provided the disk is non-viscous.
Hence, this approach enables to demonstrate that the pure hydrodynamic
turbulence\ in accretion disks is possible. It is interesting that a
turbulent magnetic disk tends to be Keplerian. This can easily be shown by
assuming that the turbulent gas tends to flow with minimal losses, i.e. to
have the Euler number as small as possible.
\end{abstract}

\section{Introduction\label{Introduction}}

We will consider the dynamics of axisymmetric accretion disk around a
compact object. A successful theory of the process in question is mainly
developed (see, e.g. \cite{Clarke and Carswell 2007}, \cite{Frank et al.
2002}, \cite{Vietri Mario 2008}, \cite{Shapiro and Teukolsky 2004}, \cite
{Shore 2007}). Extensive use is made of simple models such as vertically
isothermal disks as well as adiabatic ones. To estimate possible errors and
limits associated to these models, the exact solutions to the steady-state
MHD system were found (see Sec. \ref{VID} and Sec. \ref{Adiabatic flow})
under the assumption that the radial components of velocity and magnetic
field are negligible. We will also consider an intermediate case (see Sec.
\ref{Perfect gas}) that is between the above-listed two axisymmetric flows.
Such an approach for the modeling of thin accretion disks turns out to be
more flexible and efficient. In particular, the question of pure
hydrodynamic turbulence\ was an open question \cite{Boss 2005}, \cite{Frank
et al. 2002}, \cite{Mukhopadhyay and Chattopadhyay 2013}, \cite{Raettig et
al. 2013} until recent years. The possibility for finite disturbances to
develop turbulence in the nonlinear regime was demonstrated by O. A.
Kuznetsov in \cite{Fridman et al 2006} and in doing so he has disproved the
well-known arguments that pure hydrodynamic turbulence cannot be a
self-sustaining source of viscosity in accretion disks (see, e.g., \cite
{Frank et al. 2002} and references therein). The other possible origins of
pure hydrodynamic turbulence have been investigated in \cite{Mukhopadhyay
and Chattopadhyay 2013}, \cite{Raettig et al. 2013}. Using the approach of
Sec. \ref{Perfect gas} we will also demonstrate that the pure hydrodynamic
turbulence\ in accretion disks is possible.

The input system of MHD equations is the following (see, e.g. \cite
{Goedbloed and Poedts 2004}, \cite{Kulikovskii et al. 2001}, \cite{Landau
and Lifshitz 1984}, \cite{Loitsyanskiy 1978}, \cite{Mitchner and Kruger 1973}%
, \cite{Sedov 1971}, \cite{Shore 2007}):
\begin{equation}
\frac{\partial \rho }{\partial t}+\mathbf{\nabla \cdot }\left( \rho \mathbf{v%
}\right) =0,  \label{IS10}
\end{equation}
\begin{equation}
\frac{\partial }{\partial t}\left( \rho \mathbf{v}\right) +\mathbf{\nabla
\cdot }\left[ \rho \mathbf{vv}+\left( P+\frac{B^{2}}{8\pi }\right) \mathbf{I-%
}\frac{1}{4\pi }\mathbf{BB}\right] =\nabla \cdot \mathbf{\tau }-\rho \nabla
\Phi \mathbf{,}  \label{IS20}
\end{equation}

\begin{equation}
\frac{\partial E}{\partial t}+\mathbf{\nabla \cdot }\left[ \mathbf{v}\left(
E+P+\frac{B^{2}}{8\pi }\right) -\frac{1}{4\pi }\mathbf{B}\left( \mathbf{%
v\cdot B}\right) \right] =-\rho \mathbf{v\cdot }\nabla \Phi -\rho \dot{Q}%
-\nabla \cdot \mathbf{q,}  \label{IS25}
\end{equation}

\begin{equation}
\frac{\partial \mathbf{B}}{\partial t}=\nabla \times \left( \mathbf{v}\times
\mathbf{B}\right) ,  \label{IS30}
\end{equation}
\begin{equation}
\nabla \cdot \mathbf{B}=0,  \label{IS33}
\end{equation}
\begin{equation}
\mathbf{E}=-\frac{1}{c}\left( \mathbf{v}\times \mathbf{B}\right) ,
\label{IS40}
\end{equation}
where $\rho $, $\mathbf{v}$, $P$, $\mathbf{B}$, $\mathbf{E}$, $\mathbf{\tau }
$, $\mathbf{q}$, and $\Phi $ denote the density, velocity, pressure,
magnetic induction field, electric field, stress tensor, heat current, and
gravitational potential, respectively, $B$ $=$ $\left| \mathbf{B}\right| $, $%
v$ $=$ $\left| \mathbf{v}\right| $, $E$ $=$ $\rho e_{p}$ $+$ $0.5\rho v^{2}$
$+$ $B^{2}\diagup \left( 8\pi \right) $ denotes the total energy per unit
volume with $e_{p}$ being the internal energy per unit mass for the plasma, $%
\dot{Q}$ denotes the local cooling rate \cite[p. 142]{Clarke and Carswell
2007}. It is, mainly, assumed that
\begin{equation}
\Phi =-G\frac{M}{\sqrt{r^{2}+z^{2}}},\qquad G,M=const.  \label{IS45}
\end{equation}
The heat conductive flux, $\mathbf{q}$, may be expressed as
\begin{equation}
\mathbf{q}=-\lambda _{T}\nabla \cdot T,  \label{EM110}
\end{equation}
where $\lambda _{T}$ denotes the thermal conductivity, $T$ denotes the
temperature. The stress tensor, $\mathbf{\tau }$, is the sum of two tensors,
$\mathbf{\tau =\tau }_{v}+\mathbf{\tau }_{t}$, namely, the viscous, $\mathbf{%
\tau }_{v}$, and the turbulent, $\mathbf{\tau }_{t}$, stress tensors:
\begin{equation}
\mathbf{\tau }_{v}\approx \mu _{v}\left[ \nabla \mathbf{v}+\left( \nabla
\mathbf{v}\right) ^{\ast }\right] -\frac{2}{3}\mu _{v}\nabla \cdot \mathbf{vI%
},  \label{EM120}
\end{equation}
\begin{equation}
\mathbf{\tau }_{t}\approx \mu _{t}\left[ \nabla \mathbf{v}+\left( \nabla
\mathbf{v}\right) ^{\ast }\right] -\frac{2}{3}\left( \mu _{t}\nabla \cdot
\mathbf{v+}\rho \overline{\kappa }\right) \mathbf{I},  \label{EMT10}
\end{equation}
where $\left( \mbox{\hspace{2.0mm}}\right) ^{\ast }$ denotes a conjugate
tensor, $\mu _{v}$ denotes the dynamic viscosity, $\mu _{t}$ and $\overline{%
\kappa }$ denote the turbulent viscosity and the kinetic energy of
turbulence, respectively (see, e.g., \cite{Anderson et al. 1984}, \cite
{Fridman et al 2006} and references therein). We will also use the viscosity
$\mu =\mu _{v}+\mu _{t}$. Obviously, if the flow is laminar, then $\overline{%
\kappa }=0$ and $\mu $ is the dynamic viscosity.

In this paper, the following three axisymmetric flows in cylindrical
coordinates, $\left( r,\varphi ,z\right) $, will be considered.

1) A vertically isothermal disk, where the temperature is a pre-assigned
value,
\begin{equation}
P=\rho RT,\quad \frac{\partial T}{\partial z}=0,\qquad R=const.
\label{ISD50}
\end{equation}

2) An adiabatic disk, i.e.
\begin{equation}
P=K\rho ^{\gamma },\quad \gamma >1,\quad \gamma ,K=const.  \label{ISD60}
\end{equation}

3) An intermediate case that is between the above-listed two, in some
measure opposite, axisymmetric flows. Such an approach permits us to avoid
the solution of energy equation.

Let us note that the number densities of ions and electrons at any point are
approximately equal, and, hence, a plasma must always be close to charge
neutrality. Even a small charge imbalance would create huge electric fields
which would move the plasma particles so as to restore neutrality very
quickly \cite{Goedbloed and Poedts 2004}, \cite{Frank et al. 2002}. The
plasma maintains charge neutrality to a high degree of accuracy. However,
local charge imbalances may be produced by thermal fluctuations \cite
{Goedbloed and Poedts 2004}. To estimate their size, it can be used the
Debye length, $\lambda _{D}$, which is the typical size of a region over
which the charge imbalance may occur \cite{Clarke and Carswell 2007}, \cite
{Goedbloed and Poedts 2004}, \cite{Frank et al. 2002}: $\lambda _{D}\approx
70\sqrt{T\diagup n_{e}}$ $m$, where $n_{e}$ denotes the number density of
electrons, $T$ denotes the temperature. The length scale, $\lambda _{s}$, of
plasma dynamics should be much larger than $\lambda _{D}$. For example,
inserting the numbers for coronal plasma, we find \cite{Goedbloed and Poedts
2004} $\lambda _{D}=0.07$ $m$. Considering typical length scales of coronal
loops \cite{Goedbloed and Poedts 2004}, $\lambda _{s}=10000$ $km$, we can
see that the condition $\lambda _{s}\gg \lambda _{D}$ is easily satisfied.
We will, mainly, consider the flow at the periphery of accretion disk and,
hence, the length scale of plasma dynamics will be much larger than the
Debye length. Actually, the mean density of gas in the Milky Way is a
million per cubic metre \cite{Clarke and Carswell 2007}. Then, assuming that
the gas temperature is close to $100^{\circ }$ $K$, we find that the Debye
length will be less than $0.7$ $m$. If, however, inside the disk there is a
charge density, then it gives rise to an electric field outside the disc
which is available to pull charges out of the disc \cite{Frank et al. 2002}.
Hence, in the case of at least steady-state flow, we may write (see, e.g.,
\cite[p. 188]{Clarke and Carswell 2007}, \cite[p. 275]{Shore 2007}) that
\begin{equation}
\nabla \cdot \mathbf{E}=0\ \mathbf{\Rightarrow \ }\nabla \cdot \left(
\mathbf{v}\times \mathbf{B}\right) =0.  \label{ISD80}
\end{equation}

Let us introduce the following characteristic quantities: $t_{\ast }$, $%
l_{\ast }$, $\rho _{\ast }$, $v_{\ast }$, $p_{\ast }$, $T_{\ast }$, $\mu
_{\ast }$, $\overline{\kappa }_{\ast }$, and $B_{\ast }$ for, respectively,
time, length, density, velocity, pressure, temperature, viscosity, kinetic
energy of turbulence, and magnetic field. The following notation will also
be used:

\begin{equation}
S_{h}=\frac{l_{\ast }}{v_{\ast }t_{\ast }},\ E_{u}=\frac{p_{\ast }}{\rho
_{\ast }v_{\ast }^{2}},\ \beta =\frac{4\pi p_{\ast }}{B_{\ast }^{2}},\ F_{r}=%
\frac{v_{\ast }^{2}l_{\ast }}{GM},\ R_{e}=\frac{\rho _{\ast }v_{\ast
}l_{\ast }}{\mu _{\ast }},\ \vartheta _{ke}=\frac{2\overline{\kappa }_{\ast }%
}{3v_{\ast }^{2}},  \label{EM230}
\end{equation}
where $S_{h}$, $E_{u}$, $F_{r}$, and $R_{e}$ denote, respectively, Strouhal,
Euler, Froude, and Reynolds numbers. For axisymmetrical flow, we have, in
view of (\ref{EM230}), the following non-dimensional system of PDEs (Partial
Differential Equations).
\begin{equation}
S_{h}\frac{\partial \rho }{\partial t}+\frac{1}{r}\frac{\partial \left(
r\rho v_{r}\right) }{\partial r}+\frac{\partial \left( \rho v_{z}\right) }{%
\partial z}=0,  \label{EM240}
\end{equation}
\begin{equation*}
S_{h}\frac{\partial \rho v_{r}}{\partial t}+\frac{1}{r}\frac{\partial }{%
\partial r}r\left( \rho v_{r}^{2}-\frac{E_{u}}{\beta }B_{r}^{2}\right) +%
\frac{\partial }{\partial z}\left( \rho v_{r}v_{z}-\frac{E_{u}}{\beta }%
B_{r}B_{z}\right) +
\end{equation*}
\begin{equation*}
\frac{E_{u}}{\beta }\frac{B_{\varphi }^{2}}{r}-\frac{\rho v_{\varphi }^{2}}{r%
}=-\frac{\partial }{\partial r}\left( E_{u}P+\frac{E_{u}}{\beta }\frac{B^{2}%
}{2}\right) -\frac{\rho }{F_{r}}\frac{\partial \Phi }{\partial r}+
\end{equation*}
\begin{equation*}
\frac{1}{R_{e}}\left\{ \frac{\partial }{\partial r}\left[ 2\mu \frac{%
\partial v_{r}}{\partial r}-\frac{2}{3}\mu \left( \frac{1}{r}\frac{\partial
\left( rv_{r}\right) }{\partial r}+\frac{\partial v_{z}}{\partial z}\right) %
\right] +\right.
\end{equation*}
\begin{equation}
\left. \frac{\partial }{\partial z}\mu \left( \frac{\partial v_{r}}{\partial
z}+\frac{\partial v_{z}}{\partial r}\right) +\frac{2\mu }{r}\left( \frac{%
\partial v_{r}}{\partial r}\mathbf{-}\frac{v_{r}}{r}\right) \right\}
-\vartheta _{ke}\frac{\partial }{\partial r}\rho \overline{\kappa },
\label{EM250}
\end{equation}
\begin{equation*}
S_{h}\frac{\partial \rho v_{\varphi }}{\partial t}+\frac{1}{r}\frac{\partial
}{\partial r}r\left( \rho v_{\varphi }v_{r}-\frac{E_{u}}{\beta }B_{\varphi
}B_{r}\right) +
\end{equation*}
\begin{equation*}
\frac{\partial }{\partial z}\left( \rho v_{\varphi }v_{z}-\frac{E_{u}}{\beta
}B_{\varphi }B_{z}\right) +\frac{\rho v_{\varphi }v_{r}}{r}-\frac{E_{u}}{%
\beta }\frac{B_{\varphi }B_{r}}{r}=
\end{equation*}
\begin{equation}
\frac{1}{R_{e}}\left\{ \frac{\partial }{\partial r}\left[ \mu r\frac{%
\partial }{\partial r}\left( \frac{v_{\varphi }}{r}\right) \right] +\frac{%
\partial }{\partial z}\left( \mu \frac{\partial v_{\varphi }}{\partial z}%
\right) +2\mu \frac{\partial }{\partial r}\left( \frac{v_{\varphi }}{r}%
\right) \right\} ,  \label{EM270}
\end{equation}
\begin{equation*}
S_{h}\frac{\partial \rho v_{z}}{\partial t}+\frac{1}{r}\frac{\partial }{%
\partial r}r\left( \rho v_{z}v_{r}-\frac{E_{u}}{\beta }B_{z}B_{r}\right) +%
\frac{\partial }{\partial z}\left( \rho v_{z}^{2}-\frac{E_{u}}{\beta }%
B_{z}^{2}\right) =
\end{equation*}
\begin{equation*}
-\frac{\partial }{\partial z}\left( E_{u}P+\frac{E_{u}}{\beta }\frac{B^{2}}{2%
}\right) -\frac{\rho }{F_{r}}\frac{\partial \Phi }{\partial z}+
\end{equation*}
\begin{equation*}
\frac{1}{R_{e}}\left\{ \frac{\partial }{\partial r}\left[ \mu \left( \frac{%
\partial v_{r}}{\partial z}+\frac{\partial v_{z}}{\partial r}\right) \right]
+\frac{\partial }{\partial z}\left[ 2\mu \frac{\partial v_{z}}{\partial z}-%
\frac{2}{3}\mu \left( \frac{1}{r}\frac{\partial \left( rv_{r}\right) }{%
\partial r}+\frac{\partial v_{z}}{\partial z}\right) \right] +\right.
\end{equation*}
\begin{equation}
\left. \frac{\mu }{r}\left( \frac{\partial v_{r}}{\partial z}+\frac{\partial
v_{z}}{\partial r}\right) \right\} -\vartheta _{ke}\frac{\partial }{\partial
z}\rho \overline{\kappa },  \label{EM280}
\end{equation}
\begin{equation}
S_{h}\frac{\partial B_{r}}{\partial t}+\frac{\partial \left(
v_{z}B_{r}-v_{r}B_{z}\right) }{\partial z}=0,  \label{EM284}
\end{equation}
\begin{equation}
S_{h}\frac{\partial B_{\varphi }}{\partial t}+\frac{\partial \left(
v_{r}B_{\varphi }-v_{\varphi }B_{r}\right) }{\partial r}+\frac{\partial
\left( v_{z}B_{\varphi }-v_{\varphi }B_{z}\right) }{\partial z}=0,
\label{EM288}
\end{equation}
\begin{equation}
S_{h}\frac{\partial B_{z}}{\partial t}+\frac{1}{r}\frac{\partial r\left(
v_{r}B_{z}-v_{z}B_{r}\right) }{\partial r}=0,  \label{EM286}
\end{equation}
\begin{equation}
\frac{1}{r}\frac{\partial rB_{r}}{\partial r}+\frac{\partial B_{z}}{\partial
z}=0.  \label{EM289}
\end{equation}
It is, mainly, assumed that
\begin{equation}
\Phi =-\frac{1}{\sqrt{r^{2}+z^{2}}}.  \label{EM290}
\end{equation}
Notice, if $\overline{\kappa }=0$ and $\mu =\mu _{v}$ in (\ref{EM250})-(\ref
{EM280}), then the flow is laminar.

We will consider, in general, accretion disks. Hence, it is assumed that the
values $\rho $, $v_{r}$, $v_{\varphi }$, $P$, $\Phi $ are even functions of $%
z$, whereas $v_{z}$ is an odd one. In such a case, in view of (\ref{EM240})-(%
\ref{EM290}), there exist two possibilities: 1) $B_{z}$ is an even function
of $z$, whereas $B_{\varphi }$ and $B_{r}$ are odd ones; 2) the values $%
B_{\varphi }$ and $B_{r}$ are even functions of $z$, whereas $B_{z}$ is an
odd one. We will consider the first possibility.

If $v_{r}=v_{z}=0$, then there exists the third possibility, namely, the
values $B_{\varphi }$ and $B_{z}$ are even functions of $z$, whereas $B_{r}$
is an odd one. In such a case the magnetic field will be unstable provided $%
B_{\varphi }\neq 0$. Actually, if $v_{r}\neq 0$, then $B_{\varphi }$ will be
an odd function of $z$, since we consider the case when $B_{z}$ is an even
function of $z$. It is very important to note that the solution such that $%
B_{\varphi }$ $\left( \neq 0\right) $ is an even functions of $z$ can not be
obtained as a limiting case (namely, as $v_{r}\rightarrow 0$) of the motion
under $v_{r}\neq 0$. Hence, any solution for $B_{\varphi }$ $\left( \neq
0\right) $ such that it is not an odd function of $z$ may be seen as
unstable, as any infinitesimal variation, $\delta v_{r}$ $\left( \neq
0\right) $, gives rise to a finite response in the magnetic field.

In the case of steady-state flow, we sometimes assume that the flow is
charge-neutral, (\ref{ISD80}), i.e.
\begin{equation}
\frac{\partial r\left( v_{\varphi }B_{z}-v_{z}B_{\varphi }\right) }{%
r\partial r}+\frac{\partial \left( v_{r}B_{\varphi }-v_{\varphi
}B_{r}\right) }{\partial z}=0.  \label{EMD320}
\end{equation}

\section{Vertically isothermal disk\label{VID}}

In this section we construct steady-state solutions for the system (\ref
{EM240})-(\ref{EM289}) provided $R_{e}\rightarrow \infty $, $\vartheta
_{ke}=0$, and
\begin{equation}
B_{r}=0,\quad v_{r}=0,\quad \frac{\partial T}{\partial z}=0.  \label{S03}
\end{equation}
Let us note that the equality $v_{r}=0$ implies $v_{z}=0$. It could be
easily seen from the steady-state version of (\ref{EM240}). We take $p_{\ast
}=\rho _{\ast }RT_{\ast }$ and, hence, we obtain from (\ref{ISD50}) that
\begin{equation}
P=\rho T.  \label{VT10}
\end{equation}
The temperature in (\ref{VT10}) is assumed to be a preassigned function of $%
r $.

In such a case, the MHD system is reduced to the following.
\begin{equation}
\frac{E_{u}}{\beta }\frac{B_{\varphi }^{2}}{r}-\frac{\rho v_{\varphi }^{2}}{r%
}=-\frac{\partial }{\partial r}\left( E_{u}P+\frac{E_{u}}{\beta }\frac{B^{2}%
}{2}\right) -\frac{\rho }{F_{r}}\frac{\partial \Phi }{\partial r},
\label{S20}
\end{equation}
\begin{equation}
\frac{\partial \left( B_{\varphi }B_{z}\right) }{\partial z}=0,  \label{S30}
\end{equation}
\begin{equation}
-\frac{\partial }{\partial z}\left( \frac{E_{u}}{\beta }B_{z}^{2}\right) =-%
\frac{\partial }{\partial z}\left( E_{u}P+\frac{E_{u}}{\beta }\frac{B^{2}}{2}%
\right) -\frac{\rho }{F_{r}}\frac{\partial \Phi }{\partial z},  \label{S40}
\end{equation}
\begin{equation}
\frac{\partial \left( v_{\varphi }B_{z}\right) }{\partial z}=0,  \label{S60}
\end{equation}
\begin{equation}
\frac{\partial B_{z}}{\partial z}=0.  \label{S80}
\end{equation}

In view of (\ref{S80}), (\ref{S60}), and (\ref{S30}), we obtain:
\begin{equation}
B_{z}=B_{z}\left( r\right) ,\ v_{\varphi }=v_{\varphi }\left( r\right) ,\
B_{\varphi }=B_{\varphi }\left( r\right) .  \label{S90}
\end{equation}
Then we obtain, instead of (\ref{S20})-(\ref{S80}):
\begin{equation}
\frac{E_{u}}{\beta }\frac{B_{\varphi }^{2}}{r}-\frac{\rho v_{\varphi }^{2}}{r%
}=-\frac{\partial }{\partial r}\left( E_{u}P+\frac{E_{u}}{\beta }\frac{%
B_{z}^{2}+B_{\varphi }^{2}}{2}\right) -\frac{\rho }{F_{r}}\frac{\partial
\Phi }{\partial r},  \label{S120}
\end{equation}
\begin{equation}
0=-\frac{\partial }{\partial z}\left( E_{u}P\right) -\frac{\rho }{F_{r}}%
\frac{\partial \Phi }{\partial z}.  \label{S130}
\end{equation}
Let
\begin{equation}
C_{\rho }=\left. \rho \right| _{z=0},\quad \phi =\left. \Phi \right| _{z=0},
\label{ES07}
\end{equation}
By virtue of (\ref{VT10}), we find from (\ref{S130}) that
\begin{equation}
\rho =C_{\rho }\exp \left( \frac{\phi -\Phi }{TF_{r}E_{u}}\right) ,\quad
C_{\rho }=C_{\rho }\left( r\right) ,\ \phi =\phi \left( r\right) \equiv
\left. \Phi \right| _{z=0}.  \label{ES09}
\end{equation}
If (\ref{EM290}) is valid, then
\begin{equation}
\rho =C_{\rho }\exp \left[ \frac{1}{TF_{r}E_{u}}\left( \frac{1}{\sqrt{%
r^{2}+z^{2}}}-\frac{1}{r}\right) \right] ,\quad C_{\rho }=C_{\rho }\left(
r\right) .  \label{ES10}
\end{equation}
Let, in general,
\begin{equation}
\frac{\partial \Phi }{\partial z}\neq 0.  \label{ES70}
\end{equation}
Eq. (\ref{S120}) must be valid under all values of $z\geq 0$. After
differentiation (\ref{S120}) over $z$, in view of (\ref{S130}) we obtain
\begin{equation}
\frac{v_{\varphi }^{2}}{r}\frac{\rho }{TE_{u}}=\frac{\partial \rho }{%
\partial r}+\frac{\rho }{TF_{r}E_{u}}\frac{\partial \Phi }{\partial r}.
\label{ES80}
\end{equation}
Since, in view of (\ref{ES09}),
\begin{equation}
\frac{\partial \rho }{\partial r}=\left[ \frac{\partial C_{\rho }}{\partial r%
}+C_{\rho }\frac{\partial }{\partial r}\left( \frac{\phi -\Phi }{TF_{r}E_{u}}%
\right) \right] \exp \left( \frac{\phi -\Phi }{TF_{r}E_{u}}\right) ,
\label{ES90}
\end{equation}
we find from (\ref{ES80}) that
\begin{equation}
\frac{v_{\varphi }^{2}}{r}\frac{C_{\rho }}{TE_{u}}=\frac{\partial C_{\rho }}{%
\partial r}+\frac{C_{\rho }}{TF_{r}E_{u}}\frac{\partial \phi }{\partial r}-%
\frac{C_{\rho }\left( \phi -\Phi \right) }{T^{2}F_{r}E_{u}}\frac{\partial T}{%
\partial r}.  \label{ES130}
\end{equation}
After differentiation (\ref{ES130}) with respect to $z$, we obtain:
\begin{equation}
\frac{C_{\rho }}{T^{2}F_{r}E_{u}}\frac{\partial \Phi }{\partial z}\frac{%
\partial T}{\partial r}=0.  \label{ES135}
\end{equation}
Since $C_{\rho }\neq 0$, we obtain from (\ref{ES135}), in view of (\ref{ES70}%
), that
\begin{equation}
T=const,  \label{ES140}
\end{equation}
and, hence,
\begin{equation}
C_{\rho }=const\exp \left( \int \frac{v_{\varphi }^{2}}{rTE_{u}}dr-\frac{%
\phi }{TF_{r}E_{u}}\right) .  \label{ES190}
\end{equation}
If (\ref{EM290}) is valid, then $\phi =-1\diagup r$. Let the disk be
Keplerian, i.e.
\begin{equation}
v_{\varphi }=\frac{1}{\sqrt{rF_{r}}},  \label{ES200}
\end{equation}
then
\begin{equation}
C_{\rho }=const\exp \left( \int \frac{dr}{r^{2}TF_{r}E_{u}}+\frac{1}{%
rTF_{r}E_{u}}\right) =const.  \label{ES210}
\end{equation}
Thus, the assumption that the motion is Keplerian leads to a constant
density at the midplane. Let us consider a vortex motion, i.e.
\begin{equation}
v_{\varphi }=\frac{C_{\varphi }}{r},\quad C_{\varphi }=const,  \label{ES220}
\end{equation}
and let, for the sake of simplicity, $C_{\varphi }=1\diagup \sqrt{F_{r}}$,
then, in general, we have
\begin{equation}
\frac{\partial C_{\rho }}{\partial r}<0.  \label{ES230}
\end{equation}
The density at the midplane, in view of (\ref{ES190}), will be the
following.
\begin{equation}
C_{\rho }=const\exp \left( \frac{1}{rTF_{r}E_{u}}-\frac{1}{2r^{2}TF_{r}E_{u}}%
\right) .  \label{ES240}
\end{equation}
Substituting (\ref{ES09}), (\ref{ES140}), and (\ref{ES190}) into (\ref{S120}%
), we obtain the following equation in $B_{\varphi }$ and $B_{z}$.
\begin{equation}
\frac{B_{\varphi }^{2}}{r}+\frac{\partial }{\partial r}\left( \frac{%
B_{z}^{2}+B_{\varphi }^{2}}{2}\right) =0.  \label{ES250}
\end{equation}
We can see from (\ref{S90}) that $B_{\varphi }$ $\left( \neq 0\right) $ is
an even function of $z$, and, hence, $B_{\varphi }$ is unstable (see Sec.
\ref{Introduction}). Thus, $B_{\varphi }=0$ is a possibly stable solution,
and, by virtue of (\ref{ES250}), we find that
\begin{equation}
B_{z}=const.  \label{ES270}
\end{equation}

The semi-thickness, $H$, of disk is often (e.g. \cite{Duric 2004}) defined
as
\begin{equation}
H=\frac{1}{C_{\rho }}\int\limits_{0}^{\infty }\rho dz,\quad C_{\rho }=\left.
\rho \right| _{z=0}.  \label{DT10}
\end{equation}
Notice, using the exact solution (\ref{ES10}) in (\ref{DT10}) we find that $%
H\rightarrow \infty $ provided that $TF_{r}E_{u}\neq 0$. Thus, even if the
value of $E_{u}$ be small but finite, the disk cannot be thin in terms of (%
\ref{DT10}). Instead of the exact solution, (\ref{ES10}), it can be used the
following approximation (Cf. \cite{Nature Publishing Group 2001}, \cite
{Shore 2007}) for small values of $z$.
\begin{equation}
\rho \approx C_{\rho }\exp \left( -\frac{z^{2}}{2TF_{r}E_{u}r^{3}}\right)
\equiv C_{\rho }\exp \left[ -\frac{1}{2}\left( z\diagup H\right) ^{2}\right]
,\quad H=r\sqrt{rTF_{r}E_{u}}.  \label{DT20}
\end{equation}
Let us note that the semi-thickness $H\propto r\sqrt{Tr}$ in (\ref{DT20}).
Analogous formulae can be found in many monographes (see, e.g., \cite{Duric
2004}, \cite{Nature Publishing Group 2001}, \cite{Shore 2007} and references
therein). Thus, in the case of isothermal flow we have

\begin{equation}
H\propto r^{1.5}.  \label{DT30}
\end{equation}

Till now we did not use the assumption that the flow is electrically
neutral. Let us now assume that (\ref{EMD320}) is valid. Then, in view of (%
\ref{S03}), we find
\begin{equation}
\frac{\partial r\left( v_{\varphi }B_{z}\right) }{r\partial r}=0.
\label{ES260}
\end{equation}
Hence, in view of (\ref{ES270}), we obtain from (\ref{ES260}) that (\ref
{ES220}) is valid, i.e. we have the vortex flow.

\section{Adiabatic flow\label{Adiabatic flow}}

In this section we intend to find an exact steady-state solution to the MHD
system (\ref{EM240})-(\ref{EM289}) provided $R_{e}\rightarrow \infty $, $%
\vartheta _{ke}=0$, and
\begin{equation}
v_{r}=0,\quad B_{r}=0.  \label{AF20}
\end{equation}
Let us remind that the equality $v_{r}=0$ implies $v_{z}=0$. It could be
easily seen from the steady-state version of (\ref{EM240}). We take $p_{\ast
}=K\rho _{\ast }^{\gamma }$ and, hence, we obtain from (\ref{ISD60}) that
\begin{equation}
P=\rho ^{\gamma },\quad \gamma >1.  \label{AF10}
\end{equation}

Then, by analogy with Sec. \ref{VID}, we obtain:
\begin{equation}
B_{z}=B_{z}\left( r\right) ,\ v_{\varphi }=v_{\varphi }\left( r\right) ,\
B_{\varphi }=B_{\varphi }\left( r\right) .  \label{AF30}
\end{equation}
\begin{equation}
\frac{E_{u}}{\beta }\frac{B_{\varphi }^{2}}{r}-\frac{\rho v_{\varphi }^{2}}{r%
}=-\frac{\partial }{\partial r}\left( E_{u}P+\frac{E_{u}}{\beta }\frac{%
B_{z}^{2}+B_{\varphi }^{2}}{2}\right) -\frac{\rho }{F_{r}}\frac{\partial
\Phi }{\partial r},  \label{AF40}
\end{equation}
\begin{equation}
0=-\frac{\partial }{\partial z}\left( E_{u}P\right) -\frac{\rho }{F_{r}}%
\frac{\partial \Phi }{\partial z}.  \label{AF50}
\end{equation}

It is assumed that $P\left( r,z\right) _{z=H}=0$ and, hence, $\rho \left(
r,z\right) _{z=H}=0$, where $2H\left( r\right) $ denotes the height of disk.
Let
\begin{equation}
\Psi =\frac{\gamma }{\gamma -1}\rho ^{\gamma -1}\ \Rightarrow \ \frac{1}{%
\rho }\nabla P=\nabla \Psi .  \label{AS30}
\end{equation}
By virtue of (\ref{AS30}), we rewrite (\ref{AF50}) to read
\begin{equation}
\frac{\partial \Psi }{\partial z}+\frac{1}{F_{r}E_{u}}\frac{\partial \Phi }{%
\partial z}=0.  \label{AS35}
\end{equation}
We find from (\ref{AS35})
\begin{equation}
\Psi +\frac{\Phi }{F_{r}E_{u}}=C\left( r\right) .  \label{AS40}
\end{equation}
Since $\rho \left( r,z\right) _{z=H}=0$ and, hence, $\Psi \left( r,z\right)
_{z=H}=0$, we obtain from (\ref{AS40}) that
\begin{equation}
\Psi =\frac{1}{F_{r}E_{u}}\left( \frac{1}{\sqrt{r^{2}+z^{2}}}-\frac{1}{\sqrt{%
r^{2}+H^{2}}}\right) .  \label{AS50}
\end{equation}
Hence
\begin{equation}
\rho =\left[ \frac{\gamma -1}{\gamma F_{r}E_{u}}\left( \frac{1}{\sqrt{%
r^{2}+z^{2}}}-\frac{1}{\sqrt{r^{2}+H^{2}}}\right) \right] ^{1\diagup \left(
\gamma -1\right) },\ \left| z\right| \leq H.  \label{AS55}
\end{equation}
It is clear from (\ref{AS55}) that $\rho \rightarrow 0$ as $H\rightarrow 0$.

By virtue of (\ref{AF40}) and (\ref{AS30}), we find
\begin{equation}
\frac{E_{u}}{\beta }\frac{B_{\varphi }^{2}}{r}-\frac{\rho v_{\varphi }^{2}}{r%
}=-\rho E_{u}\frac{\partial \Psi }{\partial r}-\frac{\partial }{\partial r}%
\left( \frac{E_{u}}{\beta }\frac{B_{z}^{2}+B_{\varphi }^{2}}{2}\right) -%
\frac{\rho }{F_{r}}\frac{\partial \Phi }{\partial r}.  \label{AS60}
\end{equation}
Eq. (\ref{AS60}) must be valid under all values of $z\geq 0$. Since $\rho
\left( r,z\right) _{z=H}=0$, we obtain from (\ref{AS60}) that
\begin{equation}
\frac{B_{\varphi }^{2}}{r}=-\frac{\partial }{\partial r}\left( \frac{%
B_{z}^{2}+B_{\varphi }^{2}}{2}\right) .  \label{AS70}
\end{equation}
By virtue of (\ref{AS30}) and (\ref{AS70}) we obtain from (\ref{AS60}):
\begin{equation}
\frac{v_{\varphi }^{2}}{rE_{u}}=\frac{\partial \Psi }{\partial r}+\frac{1}{%
F_{r}E_{u}}\frac{\partial \Phi }{\partial r}\equiv \frac{\partial }{\partial
r}\left( \Psi +\frac{\Phi }{F_{r}E_{u}}\right) \equiv -\frac{1}{F_{r}E_{u}}%
\frac{\partial }{\partial r}\frac{1}{\sqrt{r^{2}+H^{2}}}.  \label{AS80}
\end{equation}
In view of (\ref{AS80}), we have
\begin{equation}
\frac{v_{\varphi }^{2}}{r}=-\frac{1}{F_{r}}\frac{\partial }{\partial r}\frac{%
1}{\sqrt{r^{2}+H^{2}}}.  \label{AS90}
\end{equation}

Let, for instance, the motion is Keplerian, i.e.
\begin{equation}
v_{\varphi }=\frac{1}{\sqrt{rF_{r}}},  \label{AS95}
\end{equation}
then, by virtue of (\ref{AS90}), we obtain that
\begin{equation}
H^{2}=0.  \label{AS100}
\end{equation}
Hence, in view of (\ref{AS55}), $\rho =0$.

We can see from (\ref{AF30}) that $B_{\varphi }$ $\left( \neq 0\right) $ is
an even function of $z$, and, hence, $B_{\varphi }$ is unstable (see Sec.
\ref{Introduction}). Thus, the possibly stable solution is: $B_{\varphi }=0$%
, and, by virtue of (\ref{AS70}), we find that $B_{z}=const$.

Assuming that the flow is electrically neutral \cite{Shore 2007}, we find
from (\ref{EMD320}) that the flow is the vortex, i.e.
\begin{equation}
v_{\varphi }=\frac{C_{\varphi }}{r}.  \label{AS110}
\end{equation}
Let, for the sake of simplicity,
\begin{equation}
C_{\varphi }^{2}=\frac{1}{F_{r}},  \label{AS120}
\end{equation}
then, by virtue of (\ref{AS90}), we obtain that
\begin{equation}
H=r\sqrt{4r^{2}-1}\Rightarrow H\approx 2r^{2}.  \label{AS130}
\end{equation}
Thus, in the case of electrically neutral adiabatic flow, we find that
\begin{equation}
H\propto r^{2}.  \label{AS135}
\end{equation}

\section{Perfect gas. Pre-assigned midplane temperature\label{Perfect gas}}

As it can be seen from Sec. \ref{VID}, the vertically isothermal disk will,
in fact, be totally isothermal under the assumption that the radial
components, $v_{r}$ and $B_{r}$, of velocity and magnetic field,
respectively, are negligible. Furthermore, the disk cannot be thin in terms
of, e.g., \cite{Duric 2004}. Adiabatic disks (see Sec. \ref{Adiabatic flow}%
), in contrast to vertically isothermal ones, are more trustworthy. However,
thin disk accretion must be highly nonadiabatic, as emphasized in \cite
{Shapiro and Teukolsky 2004}. Because of this, we will consider an
intermediate case that is between the above-listed two axisymmetric flows.

In this section we will, mainly, deal with geometrically thin disks. In the
case of, e.g., adiabatic flow (see Sec \ref{Adiabatic flow}) it will be
valid if $E_{u}\ll 1$. We take $p_{\ast }=\rho _{\ast }RT_{\ast }$ and,
hence, we obtain from (\ref{ISD50}) that
\begin{equation}
P=\rho T.  \label{PG10}
\end{equation}

We will consider symmetric disks and, hence, it is, in general, assumed that
$0\leq z\leq H$. Here $H$ denotes the disk semi-thickness. The temperature
at the midplane, i.e.
\begin{equation}
T_{0}\equiv \left. T\right| _{z=0}=T_{0}\left( r,t\right) ,  \label{PG20}
\end{equation}
is assumed to be a preassigned function of $r$ and $t$. It is also assumed
that $\rho \rightarrow 0$ implies $T\rightarrow 0$ as well as $\rho =0$ and $%
T=0$ if $\left| z\right| >H$, i.e., there is a vacuum outside the disk.

\subsection{Non-magnetic disk\label{Non-magnetic disk}}

It is assumed that
\begin{equation}
B_{r}=B_{z}=B_{\varphi }=0.  \label{NMD2}
\end{equation}
Let $v_{z}=v_{r}=0$, and let $R_{e}\rightarrow \infty $, $\vartheta _{ke}=0$%
. In such a case, the steady-state version of (\ref{EM240})-(\ref{EM290})
will be the following system of PDEs.

\begin{equation}
\frac{\rho v_{\varphi }^{2}}{r}=E_{u}\frac{\partial P}{\partial r}+\frac{%
\rho }{F_{r}}\frac{\partial \Phi }{\partial r},\quad P=\rho T,\quad
E_{u},F_{r}=const,  \label{NMD10}
\end{equation}
\begin{equation}
E_{u}\frac{\partial P}{\partial z}+\frac{\rho }{F_{r}}\frac{\partial \Phi }{%
\partial z}=0,\ \left| z\right| <H\left( r\right) ,\ \Phi \equiv -\frac{1}{%
\sqrt{r^{2}+z^{2}}}=-\frac{1}{r}+\frac{1}{2r^{3}}z^{2}+\ldots ,
\label{NMD20}
\end{equation}
where $H\left( r\right) $ denotes the free surface of disk. The boundary
conditions at the free surface are the following:
\begin{equation}
\left. \rho \right| _{\left| z\right| =H}=0,\ \left. T\right| _{\left|
z\right| =H}=0.  \label{NM20}
\end{equation}
Let us note that outside the disk the system (\ref{NMD10})-(\ref{NMD20}) is
fulfilled identically, since $\rho =0$ and $T=0$ at any point with $\left|
z\right| >H$.

The radial pressure force is usually assumed to be negligible in comparison
with gravity as well as with inertia force and, hence, the circular
velocity, $v_{\varphi }$, will be Keplerian with a great precision (see,
e.g., \cite{Clarke and Carswell 2007}, \cite{Fridman et al 2006}, \cite
{Hartmann 2009}, \cite{Vietri Mario 2008}). Let us consider this assertion
in more detail. Since Eqs. (\ref{NMD10})-(\ref{NMD20}) are written in a
non-dimensional form, the assumption that gravitational and centrifugal
forces in (\ref{NMD10}) dominate is equivalent to the assumption that $%
E_{u}\ll 1\diagup F_{r}$ and $E_{u}\ll 1$. The degenerate equation
corresponding to (\ref{NMD20}), i.e., the equation obtained from (\ref{NMD20}%
) by putting $E_{u}=0$, has the following solution: $\rho =0$ for all $z\neq
0$ and $\left. \rho \right| _{z=0}\geq 0$, i.e. $\left. \rho \right| _{z=0}$
can be equal to an arbitrary non-negative real number. We have from Eq. (\ref
{NMD20}) that
\begin{equation}
\left. E_{u}\frac{\partial P}{\partial z}\right| _{z=0}=0,\quad \forall
E_{u}\geq 0.  \label{NMD30}
\end{equation}
From the above it follows that the function $P=P\left( z\right) $ can have a
weak discontinuity at $z=0$ provided $E_{u}=0$. Hence, $P\left( z\right) $
is continuous for all $z$ and, hence, $P\left( z\right) \equiv 0$ since $%
\rho =0$ for all $z\neq 0$. We emphasize that with $\left. \rho \right|
_{z=0}>0$ and $\left. P\right| _{z=0}=0$ the temperature $\left. T\right|
_{z=0}=0$. Let $E_{u}>0$. In such a case, the exact solution to Eq. (\ref
{NMD20}) is the following.
\begin{equation}
\rho =\frac{\rho _{0}T_{0}}{T}\exp \left( -\int\limits_{0}^{z}\frac{1}{%
E_{u}F_{r}T}\frac{\partial \Phi }{\partial z}dz\right) ,\quad 0\leq z\leq H,
\label{NMD40}
\end{equation}
where $\rho _{0}=\left. \rho \right| _{z=0}$, $T_{0}=\left. T\right| _{z=0}$%
. Because of the symmetry, it is assumed in (\ref{NMD40}) that $0\leq z\leq
H\left( r\right) $ instead of $\left| z\right| \leq H\left( r\right) $. We
obtain from (\ref{NMD40}) that $\rho \left( z\right) \rightarrow 0$ as $%
E_{u}\rightarrow 0$ provided that $z>0$. Hence, $H\rightarrow 0$ as $%
E_{u}\rightarrow 0$. Thus, the disk will be geometrically thin if $E_{u}\ll
1\diagup F_{r}$ and $E_{u}\ll 1$. We may, also, conclude from the above that
$\left. T\right| _{z=0}\rightarrow 0$ as $E_{u}\rightarrow 0$ even if $%
\left. \rho \right| _{z=0}>0$ under $E_{u}=0$. Thus, if $E_{u}\rightarrow 0$%
, then we obtain an infinitely thin disk with $\left. \rho \right| _{z=0}=0$
or $\left. \rho \right| _{z=0}$ can be equal to an arbitrary positive real
number. If the latter is the case, the disk would consist of non-interacting
particles. As an illustration we refer to a cold dust disk consisting of
small non-interacting grains. Since the disk is gaseous, $H\rightarrow 0$ as
$E_{u}\rightarrow 0$, and in view of (\ref{NM20}), we may assume that
\begin{equation}
\rho _{0}\equiv \left. \rho \right| _{z=0}\ \underset{E_{u}\rightarrow 0}{%
\longrightarrow }\ 0.  \label{NM25}
\end{equation}
An additional justification of (\ref{NM25}) will be done in what follows.

Since the values $\rho $, $v_{\varphi }$, and $T$ are even functions of $z$,
we will use the following asymptotic expansion in the limit $z\diagup
r\rightarrow 0$:
\begin{equation}
\rho =\rho _{0}+\rho _{2}z^{2}+\ldots ,\ T=T_{0}+T_{2}z^{2}+\ldots ,\
v_{\varphi }=v_{\varphi 0}+v_{\varphi 2}z^{2}+\ldots ,  \label{NM10}
\end{equation}
where the coefficients $\rho _{0}$, $\rho _{2}$, $T_{0}$, $T_{2}$, $%
v_{\varphi 0}$, and $v_{\varphi 2}$ depend on $r$, only. Let us note that
namely $z\diagup r\ll 1$ and, hence, the power series in (\ref{NM10}) must
be represented in $z\diagup r$. However, for simplicity sake, the
denominators ($r^{i}$, $i=0,2,\ldots $) are, mainly, included into the
coefficients.

By virtue of (\ref{NMD20})-(\ref{NM20}), we obtain the following equalities:
\begin{equation}
T_{0}\rho _{2}+T_{2}\rho _{0}+\frac{\rho _{0}}{2E_{u}F_{r}r^{3}}=0,
\label{NM30}
\end{equation}
\begin{equation}
T_{0}+T_{2}H^{2}\approx 0,  \label{NM40}
\end{equation}
\begin{equation}
\rho _{0}+\rho _{2}H^{2}\approx 0.  \label{NM50}
\end{equation}
Solving the system (\ref{NM30})-(\ref{NM50}), we find that
\begin{equation}
H^{2}\approx 4T_{0}E_{u}F_{r}r^{3},  \label{NM60}
\end{equation}
\begin{equation}
T=T_{0}-\frac{1}{4E_{u}F_{r}r^{3}}z^{2}+O\left( z^{4}\right) \equiv T_{0}-%
\frac{1}{4E_{u}F_{r}r}\left( \frac{z}{r}\right) ^{2}+O\left( \left( z\diagup
r\right) ^{4}\right) ,  \label{NM70}
\end{equation}
\begin{equation}
\rho =\rho _{0}\left( 1-\frac{z^{2}}{H^{2}}\right) +O\left( z^{4}\right)
\equiv \rho _{0}-\frac{\rho _{0}}{4T_{0}E_{u}F_{r}r}\left( \frac{z}{r}%
\right) ^{2}+O\left( \left( z\diagup r\right) ^{4}\right) .  \label{NM80}
\end{equation}
By virtue of (\ref{NM10}) and (\ref{NMD10}), we find the following
differential equation in the function $\rho _{0}=\rho _{0}\left( r\right) $.
\begin{equation}
\frac{\rho _{0}v_{\varphi 0}^{2}}{r}=E_{u}\frac{\partial T_{0}\rho _{0}}{%
\partial r}+\frac{\rho _{0}}{r^{2}F_{r}},\ r>r_{0},  \label{NM90}
\end{equation}
where $T_{0}=T_{0}\left( r\right) $ is a preassigned function. We will use
the following notation $T_{0}^{0}=T_{0}\left( r_{0}\right) $. The boundary
condition is the following
\begin{equation}
\rho _{0}^{0}=\rho _{0}\left( r_{0}\right) .  \label{NM100}
\end{equation}

The degenerate equation corresponding to (\ref{NM90}) has the following
solution.

1) $\rho _{0}=0$ and the midplane circular velocity $v_{\varphi
0}=v_{\varphi 0}\left( r\right) $ is an arbitrary function.

2) $\rho _{0}=\rho _{0}\left( r\right) $ is an arbitrary function such that $%
\rho _{0}>0$ and the motion at the midplane is Keplerian:
\begin{equation}
v_{\varphi 0}=\pm \frac{1}{\sqrt{rF_{r}}}.  \label{NM110}
\end{equation}
If the latter is the case, i.e. $\rho _{0}>0$, then the disk is infinitely
thin, consists of non-interacting particles and, hence, it is natural that
all particles are in Keplerian motion.

Let $E_{u}>0$, then the solution to (\ref{NM90}), (\ref{NM100}) is the
following.
\begin{equation}
\rho _{0}=\frac{\rho _{0}^{0}T_{0}^{0}}{T_{0}}\exp \int\limits_{r_{0}}^{r}%
\frac{1}{rE_{u}T_{0}}\left( v_{\varphi 0}^{2}-\frac{1}{rF_{r}}\right) dr.
\label{NM120}
\end{equation}
Notice, be the disk Keplerian, we would obtain, by virtue of (\ref{NM120}),
that
\begin{equation}
\rho _{0}=\frac{\rho _{0}^{0}T_{0}^{0}}{T_{0}\left( r\right) },\quad \rho
_{0}^{0},\,T_{0}^{0}=const.  \label{NM130}
\end{equation}
Thus, if $\rho _{0}\left( r\right) \rightarrow 0$ as $r\rightarrow \infty $,
then, in view of (\ref{NM130}), $T_{0}\left( r\right) \rightarrow \infty $
as $r\rightarrow \infty $, and vice versa, i.e. if $T_{0}\left( r\right)
\rightarrow 0$, then $\rho _{0}\left( r\right) \rightarrow \infty $ as $%
r\rightarrow \infty $. Assuming that $T_{0}=T_{0}^{0}=const$, we obtain that
$\rho _{0}=\rho _{0}^{0}=const$ too. Hence, the assumption that the motion
is Keplerian leads to improbable density and temperature distributions at
the midplane under $r\geq r_{0}$. To avoid such unlikely solutions, we have
to accept that the motion is not Keplerian and, hence
\begin{equation}
v_{\varphi 0}^{2}<\frac{1}{rF_{r}},\quad r\rightarrow \infty .  \label{NM140}
\end{equation}

It is well known (see, e.g., \cite{Clarke and Carswell 2007}, \cite{Shapiro
and Teukolsky 2004}, \cite{Vietri Mario 2008}) that a gaseous disk formed
around a central star is in an almost Keplerian rotation with a small inward
drift velocity. Let us investigate this ``almost Keplerian'' rotation. Let $%
v_{r}\neq 0$, then, instead of (\ref{NMD10})-(\ref{NMD20}), we should
consider the following system.
\begin{equation}
\frac{1}{r}\frac{\partial \left( r\rho v_{r}\right) }{\partial r}+\frac{%
\partial \left( \rho v_{z}\right) }{\partial z}=0,  \label{NM150}
\end{equation}
\begin{equation}
\frac{1}{r}\frac{\partial }{\partial r}r\left( \rho v_{r}^{2}\right) +\frac{%
\partial }{\partial z}\left( \rho v_{r}v_{z}\right) -\frac{\rho v_{\varphi
}^{2}}{r}=-\frac{\partial }{\partial r}\left( E_{u}P\right) -\frac{\rho }{%
F_{r}}\frac{\partial \Phi }{\partial r},  \label{NM160}
\end{equation}
\begin{equation}
\frac{1}{r}\frac{\partial }{\partial r}r\left( \rho v_{\varphi }v_{r}\right)
+\frac{\partial }{\partial z}\left( \rho v_{\varphi }v_{z}\right) +\frac{%
\rho v_{\varphi }v_{r}}{r}=0,  \label{NM170}
\end{equation}
\begin{equation}
\frac{1}{r}\frac{\partial }{\partial r}r\left( \rho v_{z}v_{r}\right) +\frac{%
\partial }{\partial z}\left( \rho v_{z}^{2}\right) =-\frac{\partial }{%
\partial z}\left( E_{u}P\right) -\frac{\rho }{F_{r}}\frac{\partial \Phi }{%
\partial z}.  \label{NM180}
\end{equation}
For the sake of simplicity, we take the characteristic quantity, $l_{\ast }$%
, for length such that
\begin{equation}
r_{0}=1.  \label{NM185}
\end{equation}
Since $v_{r}$ is an even function of $z$, but $v_{z}$ is an odd one, we can
write for small values of $z$:
\begin{equation}
v_{r}=v_{r0}+v_{r2}z^{2}+\ldots ,\ v_{z}=v_{z1}z+v_{z3}z^{3}+\ldots ,
\label{NM190}
\end{equation}
where the coefficients depend on $r$, only. By virtue of (\ref{NM10}), (\ref
{NM190}), we obtain from (\ref{NM150}), (\ref{NM170}) that
\begin{equation}
\frac{1}{r}\frac{\partial \left( r\rho _{0}v_{r0}\right) }{\partial r}+\rho
_{0}v_{z1}=0,  \label{NM200}
\end{equation}
\begin{equation}
\frac{1}{r}\frac{\partial }{\partial r}r\left( \rho _{0}v_{\varphi
0}v_{r0}\right) +\rho _{0}v_{\varphi 0}v_{z1}+\frac{\rho _{0}v_{\varphi
0}v_{r0}}{r}=0.  \label{NM210}
\end{equation}
By virtue of (\ref{NM200})-(\ref{NM210}), we obtain:
\begin{equation}
\rho _{0}v_{r0}\frac{\partial rv_{\varphi 0}}{r\partial r}=0.  \label{NM215}
\end{equation}
Thus, we find the following exact solution for the midplane value, $%
v_{\varphi 0}$, of circular velocity:

\begin{equation}
v_{\varphi 0}=\frac{C_{\varphi 0}}{r},\quad C_{\varphi 0}=const.
\label{NM220}
\end{equation}

For the sake of convenience, let us represent $C_{\varphi 0}$ as a function
of the Froude number, $F_{r}$. Let the point $r=r_{m}$ be the only point
where the motion is Keplerian as well as the vortex, (\ref{NM220}). Then
\begin{equation}
\frac{C_{\varphi 0}}{r_{m}}=\frac{1}{\sqrt{r_{m}F_{r}}}\Rightarrow
C_{\varphi 0}=\sqrt{\frac{r_{m}}{F_{r}}}.  \label{CDM80}
\end{equation}
Hence,
\begin{equation}
v_{\varphi 0}\equiv \left. v_{\varphi }\right| _{z=0}=\frac{\sqrt{r_{m}}}{r%
\sqrt{F_{r}}},\quad r_{m}=const.  \label{CDM90}
\end{equation}
Notice, in the case of a non-viscous flow with $v_{r0}\neq 0$ we have the
only solution, (\ref{NM220}), for the midplane circular velocity, $%
v_{\varphi 0}$. If, however, $v_{r0}=0$, then we have infinitely many
solutions for $v_{\varphi 0}$. The only limitations are the boundary
conditions. In particular, the function $v_{\varphi 0}=v_{\varphi 0}\left(
r\right) $ must be such that $\rho _{0}\left( r\right) \rightarrow 0$ as $%
r\rightarrow \infty $ in (\ref{NM120}). Thus, if $v_{r0}=0$, then we have
the ill-posed problem, i.e. the problem is not well-posed in the sense of
Hadamard \cite{Anderson et al. 1984}. It is important to note that any
solution (excluding the vortex) for $v_{\varphi 0}$ can not be obtained as a
limiting case (namely, as $v_{r0}\rightarrow 0$) of the motion under $%
v_{r0}\neq 0$. Hence, any solution that does not coincide with the vortex
will be unstable, as any infinitesimal variation, $\delta v_{r0}\neq 0$,
gives rise to a finite response in the gas flow. The stability of the
solution (\ref{CDM90}) should be investigated. Some stability aspects of the
vortex flow will be discussed in Sec. \ref{Stability}. It is necessary to
stress that, by virtue of the solution (\ref{NM220}), we obtain from (\ref
{NM120}) the validity of (\ref{NM25}).

Thus, in contrast to the widely known assertion that the circular velocity
will be Keplerian with a great precision (see, e.g., \cite{Clarke and
Carswell 2007}, \cite{Fridman et al 2006}, \cite{Hartmann 2009}, \cite
{Vietri Mario 2008}), the forgoing proves that in the case of steady-state
non-viscous disk we obtain the vortex velocity distribution (\ref{NM220})
even if the radial pressure force is negligible in comparison with gravity
and inertia forces. The solutions similar to (\ref{NM220}) are used to
describe a large variety of flow patterns, in particular, to model cosmic
swirling jets that develop near accretion disks \cite{Shtern at el. 1997}.

Let us note that the power-law model \cite[p. 374]{Vietri Mario 2008}
\begin{equation}
v_{\varphi 0}=\frac{C_{\varphi }}{r^{\varkappa }},\quad \varkappa
,C_{\varphi }=const,  \label{CDM100}
\end{equation}
is assumed to be stable (Rayleigh stable \cite{Frank et al. 2002}) to pure
hydrodynamic perturbations under$\ 0.5\leq \varkappa <1$, since it satisfies
the following necessary and sufficient condition for the so-called Rayleigh
stability \cite[p. 78]{Beskin et al 2002}.
\begin{equation}
f_{R}^{2}\equiv \frac{1}{r^{3}}\frac{\partial }{\partial r}\left(
r^{2}\Omega \right) ^{2}>0,  \label{CDM110}
\end{equation}
where $\Omega $ denotes the angular velocity, $f_{R}^{2}$ denotes the
Rayleigh frequency. Such an assertion contradicts to the forgoing prove that
any solution that does not coincide with the vortex velocity distribution (%
\ref{NM220}) will be unstable. It should be stressed that the Rayleigh
criterion, (\ref{CDM110}), is not strictly applicable to a nebular accretion
disk \cite{Boss 2005}. This remark, \cite{Boss 2005}, is correct because the
Rayleigh criterion, (\ref{CDM110}), is justified for subsonic flows \cite
{Beskin et al 2002}, whereas we consider highly supersonic flows. Actually,
since $E_{u}\ll 1$ the Mach number $M_{s}=1\diagup \sqrt{E_{u}}\gg 1$.

Let us estimate $v_{\varphi 2}$ in (\ref{NM10}). We will use the following
asymptotic expansion
\begin{equation}
T_{0}\left( r\right) \approx \sum\limits_{n=0}^{\infty }a_{n}r^{-n}
\label{ASE10}
\end{equation}
in the limit $r\rightarrow \infty $. As usually, the series in (\ref{ASE10})
may converge or diverge, but its partial sums are good approximations to $%
T_{0}\left( r\right) $ for large enough $r$. Assuming that $T_{0}\left(
r\right) \rightarrow 0$ as $r\rightarrow \infty $, we find that $a_{0}=0$.
Then, assuming the characteristic quantity, $T_{\ast }$, for temperature
such that $a_{1}=1$, we represent $T_{0}\left( r\right) $\ in the following
form:
\begin{equation}
T_{0}\left( r\right) =\frac{1}{r}+O\left( r^{-2}\right) ,\quad r\rightarrow
\infty .  \label{ASE20}
\end{equation}
Thus, we can use the following approximation:
\begin{equation}
T_{0}\left( r\right) \approx \frac{1}{r},\quad r\rightarrow \infty .
\label{ASE25}
\end{equation}
One can easily see, for instance, that the midplane temperature of the
adiabatic disks considered in Sec. \ref{Adiabatic flow} can be approximated
by (\ref{ASE25}).

By virtue of (\ref{ASE25}), we find from (\ref{NM60}) that
\begin{equation}
H\propto r.  \label{ASE27}
\end{equation}
Let $v_{z}$ is a linear function of $z$, i.e. $v_{z}=v_{z1}z$ with $%
v_{z1}=v_{z1}\left( r\right) $. From the kinematic boundary condition at the
free surface, i.e.
\begin{equation}
\left. v_{z}\right| _{z=H}=\frac{\partial H}{\partial r}\left. v_{r}\right|
_{z=H},  \label{ASE30}
\end{equation}
we find, by virtue of (\ref{ASE25}) and (\ref{NM60}), that
\begin{equation}
v_{z1}H\approx \frac{\partial H}{\partial r}v_{r0}\Rightarrow v_{z1}\approx
\frac{v_{r0}}{r}.  \label{NM270}
\end{equation}
By virtue of (\ref{NM190}) and (\ref{NM10}), we obtain from (\ref{NM150}), (%
\ref{NM170}) that
\begin{equation}
v_{r0}\frac{\partial v_{\varphi 2}}{\partial r}+2v_{z1}v_{\varphi 2}+\frac{%
v_{\varphi 2}v_{r0}}{r}=0.  \label{NM275}
\end{equation}
Thus, by virtue of (\ref{NM270}), we find
\begin{equation}
v_{\varphi 2}\approx \frac{const}{r^{3}}.  \label{NM280}
\end{equation}
By virtue of (\ref{NM10}), (\ref{NM60}), (\ref{ASE25}), and (\ref{NM280}),
we find that
\begin{equation}
\left. v_{\varphi }\right| _{z=H}\approx \frac{C_{\varphi 0}}{r}\left(
1-C_{\varphi 2}E_{u}F_{r}\right) ,\quad C_{\varphi 2}=const>0.  \label{NM290}
\end{equation}
Since $E_{u}\ll 1$, it can be assumed that
\begin{equation}
v_{\varphi }\approx \frac{C_{\varphi 0}}{r},\quad 0\leq z\leq H.
\label{NM300}
\end{equation}
It is easy to see from (\ref{NM300}) that the Rayleigh frequency $%
f_{R}^{2}=0 $ and, hence, the vortex is Rayleigh unstable.

Let us now consider the stability problem of the power-law model from
another point of view.

\subsubsection{Instability of non-magnetic disk. Power-law model\label%
{Stability}}

Let $\overline{\varsigma }$ denote the value of a dependent variable, $%
\varsigma $, for the case of a steady-state solution. In view of (\ref
{CDM100}), (\ref{NM120}), (\ref{NM140}) and (\ref{NM185}), the solution to
the system (\ref{NMD10})-(\ref{NM20}) can be written in the form:
\begin{equation}
\overline{v}_{z}=\overline{v}_{r}=0,\ \overline{v}_{\varphi 0}\equiv \left.
\overline{v}_{\varphi }\right| _{z=0}=\frac{C_{\varphi }}{r^{\varkappa }}%
,\quad \varkappa ,C_{\varphi }=const,\ 0.5<\varkappa \leq 1,  \label{LI10}
\end{equation}
\begin{equation}
\overline{H}^{2}\approx 4\overline{T}_{0}E_{u}F_{r}r^{3},\ \overline{\rho }%
_{0}=\overline{\rho }_{0}^{0}\frac{\overline{T}_{0}^{0}}{\overline{T}_{0}}%
\exp \int\limits_{1}^{r}\frac{1}{rE_{u}\overline{T}_{0}}\left( \overline{v}%
_{\varphi 0}^{2}-\frac{1}{rF_{r}}\right) dr,  \label{LI20}
\end{equation}
where $\overline{\rho }_{0}^{0}=\left. \overline{\rho }_{0}\right| _{r=1}$, $%
\overline{T}_{0}^{0}=\left. \overline{T}_{0}\right| _{r=1}$. To represent $%
C_{\varphi }$ in (\ref{LI10})\ as a function of the Froude number, $F_{r}$,
we assume that the point $r=r_{m}$ will be the only point where the motion
is Keplerian as well as the power-law model, (\ref{LI10}). Hence
\begin{equation}
\frac{C_{\varphi }}{r_{m}^{\varkappa }}=\frac{1}{\sqrt{r_{m}F_{r}}}%
\Rightarrow C_{\varphi }=\frac{r_{m}^{\varkappa -0.5}}{\sqrt{F_{r}}}.
\label{LI30}
\end{equation}

The non-linear system in perturbations $\widetilde{\varsigma }$ ($\varsigma =%
\overline{\varsigma }+\widetilde{\varsigma }$) for (\ref{EM240})-(\ref{EM280}%
) will be, under $R_{e}\rightarrow \infty $ and $\vartheta _{ke}=0$, as
follows.
\begin{equation}
S_{h}\frac{\partial \left( \overline{\rho }_{0}+\widetilde{\rho }_{0}\right)
}{\partial t}+\frac{1}{r}\frac{\partial r\left( \overline{\rho }_{0}+%
\widetilde{\rho }_{0}\right) \widetilde{v}_{r0}}{\partial r}+\left(
\overline{\rho }_{0}+\widetilde{\rho }_{0}\right) \widetilde{v}_{z1}=0,
\label{LI40}
\end{equation}
\begin{equation*}
S_{h}\frac{\partial \left( \overline{\rho }_{0}+\widetilde{\rho }_{0}\right)
\widetilde{v}_{r0}}{\partial t}+\frac{1}{r}\frac{\partial }{\partial r}%
r\left( \overline{\rho }_{0}+\widetilde{\rho }_{0}\right) \widetilde{v}%
_{r0}^{2}+
\end{equation*}
\begin{equation*}
\left( \overline{\rho }_{0}+\widetilde{\rho }_{0}\right) \widetilde{v}_{r0}%
\widetilde{v}_{z1}-\frac{\left( \overline{\rho }_{0}+\widetilde{\rho }%
_{0}\right) \left( \overline{v}_{\varphi 0}+\widetilde{v}_{\varphi 0}\right)
^{2}}{r}=
\end{equation*}
\begin{equation}
-E_{u}\frac{\partial }{\partial r}\left( \overline{\rho }_{0}+\widetilde{%
\rho }_{0}\right) \left( \overline{T}_{0}+\widetilde{T}_{0}\right) -\frac{%
\overline{\rho }_{0}+\widetilde{\rho }_{0}}{r^{2}F_{r}},  \label{LI50}
\end{equation}
\begin{equation*}
S_{h}\frac{\partial \left( \overline{\rho }_{0}+\widetilde{\rho }_{0}\right)
\left( \overline{v}_{\varphi 0}+\widetilde{v}_{\varphi 0}\right) }{\partial t%
}+\frac{1}{r}\frac{\partial }{\partial r}r\left( \overline{\rho }_{0}+%
\widetilde{\rho }_{0}\right) \left( \overline{v}_{\varphi 0}+\widetilde{v}%
_{\varphi 0}\right) \widetilde{v}_{r0}+
\end{equation*}
\begin{equation}
\left( \overline{\rho }_{0}+\widetilde{\rho }_{0}\right) \left( \overline{v}%
_{\varphi 0}+\widetilde{v}_{\varphi 0}\right) \widetilde{v}_{z1}+\frac{%
\left( \overline{\rho }_{0}+\widetilde{\rho }_{0}\right) \left( \overline{v}%
_{\varphi 0}+\widetilde{v}_{\varphi 0}\right) \widetilde{v}_{r0}}{r}=0,
\label{LI60}
\end{equation}
\begin{equation*}
S_{h}\frac{\partial \left( \overline{\rho }_{0}+\widetilde{\rho }_{0}\right)
\widetilde{v}_{z1}}{\partial t}+\frac{1}{r}\frac{\partial }{\partial r}%
r\left( \overline{\rho }_{0}+\widetilde{\rho }_{0}\right) \widetilde{v}_{z1}%
\widetilde{v}_{r0}+2\left( \overline{\rho }_{0}+\widetilde{\rho }_{0}\right)
\widetilde{v}_{z1}^{2}=
\end{equation*}
\begin{equation}
-2E_{u}\frac{\partial }{\partial z}\left[ \left( \overline{\rho }_{0}+%
\widetilde{\rho }_{0}\right) \left( \overline{T}_{2}+\widetilde{T}%
_{2}\right) +\left( \rho _{2}+\widetilde{\rho }_{2}\right) \left( \overline{T%
}_{0}+\widetilde{T}_{0}\right) \right] -\frac{\overline{\rho }_{0}+%
\widetilde{\rho }_{0}}{F_{r}r^{3}}.  \label{LI70}
\end{equation}

With the basic Parker's assumption \cite{Parker 1974}, as applied to
perturbations, $\widetilde{\varsigma }$, we may assume that, within the
disk, radial derivatives, $\partial \widetilde{\varsigma }\diagup \partial r$%
, are negligible in comparison to vertical ones. Using the Parker's
assumption \cite{Parker 1974} we can reduce (\ref{LI40})-(\ref{LI70}) to an
ODE (ordinary differential equation) system for a subsequent stability
investigation. In this connection \ we note that the following question
remains to be answered. What a class of admissible functions, $\widetilde{%
\varsigma }_{i}=\widetilde{\varsigma }_{i}\left( r,t\right) $, in the
expansions
\begin{equation}
\widetilde{\varsigma }=\sum_{i=0}^{\infty }\widetilde{\varsigma }_{i}z^{i},
\label{LI75}
\end{equation}
should be taken into account? Obviously, it must be such that, at least, the
main terms of expansions may be dropped under the derivative over the
coordinate $r$. Assuming $\partial \widetilde{\varsigma }_{0}\diagup
\partial r=0$ in (\ref{LI40})-(\ref{LI70}) we find that $\widetilde{%
\varsigma }_{0}\equiv 0$, since $\left. \widetilde{\varsigma }\right|
_{r\rightarrow \infty }=0$. In order to circumvent this problem, we will
consider a function as admissible if it will be a singular function \cite
{Kolmogorov and Fomin 1970}. That is, this function is continous on $1\leq
r<\infty $ and the derivative over the coordinate $r$ exists and is zero
almost everywhere. It can be also assumed that the function is strictly
monotone decreasing, e.g. $\left. \widetilde{v}_{r0}\right| _{r\rightarrow
\infty }\rightarrow 0$. Cantor staircase \cite{Kolmogorov and Fomin 1970}
and Lebesgue singular function \cite{Kawamura 2010}, \cite{Kawamura 2011}
are well known examples of such functions. Singular functions occur in
physics, dynamical systems, etc. (see e.g. the references in \cite{Kawamura
2010}).

We restrict ourself to the case of linear system. Assuming the non-linear
terms, as well as $\widetilde{T}_{0}$ and $\widetilde{T}_{2}$, in (\ref{LI40}%
)-(\ref{LI70}) as negligible, we get:
\begin{equation}
S_{h}\frac{\partial \widetilde{\rho }_{0}}{\partial t}+\frac{1}{r}\frac{%
\partial r\overline{\rho }_{0}\widetilde{v}_{r0}}{\partial r}+\overline{\rho
}_{0}\widetilde{v}_{z1}=0,  \label{LI80}
\end{equation}
\begin{equation}
S_{h}\overline{\rho }_{0}\frac{\partial \widetilde{v}_{r0}}{\partial t}-%
\frac{2\overline{\rho }_{0}\overline{v}_{\varphi 0}\widetilde{v}_{\varphi 0}+%
\widetilde{\rho }_{0}\overline{v}_{\varphi 0}^{2}}{r}=-E_{u}\frac{\partial }{%
\partial r}\left( \widetilde{\rho }_{0}\overline{T}_{0}\right) -\frac{%
\widetilde{\rho }_{0}}{r^{2}F_{r}},  \label{LI90}
\end{equation}
\begin{equation}
S_{h}\frac{\partial \widetilde{v}_{\varphi 0}}{\partial t}=\left( \varkappa
-1\right) \frac{C_{\varphi }\widetilde{v}_{r0}}{r^{\varkappa +1}},
\label{LI100}
\end{equation}
\begin{equation}
S_{h}\overline{\rho }_{0}\frac{\partial \widetilde{v}_{z1}}{\partial t}%
=-2E_{u}\left( \widetilde{\rho }_{0}\overline{T}_{2}+\widetilde{\rho }_{2}%
\overline{T}_{0}\right) -\frac{\widetilde{\rho }_{0}}{r^{3}F_{r}}.
\label{LI110}
\end{equation}
It easy to see from (\ref{NM10}),\ (\ref{NM60})-(\ref{NM80}) that
\begin{equation}
\overline{T}_{2}=-\frac{1}{4E_{u}F_{r}r^{3}},\ \overline{\rho }_{2}=-\frac{%
\overline{\rho }_{0}}{4\overline{T}_{0}E_{u}F_{r}r^{3}},\ \widetilde{\rho }%
_{2}=-\frac{\widetilde{\rho }_{0}}{4\overline{T}_{0}E_{u}F_{r}r^{3}}.
\label{LI120}
\end{equation}
By virtue of (\ref{LI120}), we rewrite (\ref{LI80})-(\ref{LI110}) to read:
\begin{equation}
S_{h}\frac{\partial \widetilde{\rho }_{0}}{\partial t}=A_{12}\widetilde{v}%
_{r0}+A_{14}\widetilde{v}_{z1}=0,  \label{LI130}
\end{equation}
\begin{equation}
S_{h}\frac{\partial \widetilde{v}_{r0}}{\partial t}=A_{21}\widetilde{\rho }%
_{0}+A_{23}\widetilde{v}_{\varphi 0},  \label{LI140}
\end{equation}
\begin{equation}
S_{h}\frac{\partial \widetilde{v}_{\varphi 0}}{\partial t}=A_{32}\widetilde{v%
}_{r0},  \label{LI150}
\end{equation}
\begin{equation}
S_{h}\frac{\partial \widetilde{v}_{z1}}{\partial t}=0,  \label{LI160}
\end{equation}
where
\begin{equation*}
A_{12}=-\frac{\partial r\overline{\rho }_{0}}{r\partial r},\ A_{14}=-%
\overline{\rho }_{0},\ A_{21}=\frac{C_{\varphi }^{2}}{\overline{\rho }%
_{0}r^{2\varkappa +1}}-E_{u}\frac{\partial \overline{T}_{0}}{\overline{\rho }%
_{0}\partial r}-\frac{1}{\overline{\rho }_{0}r^{2}F_{r}},
\end{equation*}
\begin{equation}
A_{23}=\frac{2C_{\varphi }}{r^{\varkappa +1}},\quad A_{32}=\frac{\left(
\varkappa -1\right) C_{\varphi }}{r^{\varkappa +1}}.  \label{LI170}
\end{equation}
We will use the following notation: $\mathbf{x}=\left\{
x_{1},x_{2},x_{3},x_{4}\right\} ^{\ast }=\left\{ \widetilde{\rho }_{0},%
\widetilde{v}_{r0},\widetilde{v}_{\varphi 0},\widetilde{v}_{z1}\right\}
^{\ast }$. Let $t\rightarrow S_{h}t$, then the system (\ref{LI130})-(\ref
{LI160}) can be rewritten to read:
\begin{equation}
\frac{d\mathbf{x}}{dt}=\mathbf{A}\cdot \mathbf{x,}  \label{LI180}
\end{equation}
where
\begin{equation}
\mathbf{A}=\left\{
\begin{array}{cccc}
0 & A_{12} & 0 & A_{14} \\
A_{21} & 0 & A_{23} & 0 \\
0 & A_{32} & 0 & 0 \\
0 & 0 & 0 & 0
\end{array}
\right\} .  \label{LI190}
\end{equation}
The eigenvalues of $\mathbf{A}$ are the following:
\begin{equation}
\lambda _{1}=\lambda _{2}=0,\quad \lambda _{3,4}=\pm \sqrt{%
A_{23}A_{32}+A_{12}A_{21}}.  \label{LI200}
\end{equation}
Inasmuch as the eigenvalue $\lambda =0$ has algebraic multiplicity $m=2$ and
rank$\mathbf{A}=3$ provided $0.5<\varkappa <1$, the solution to (\ref{LI180}%
) contains secular terms, namely the terms proportional to the time. Hence,
the solution, (\ref{LI10})-(\ref{LI20}) under $0.5<\varkappa <1$, to the
system (\ref{NMD10})-(\ref{NMD20}) is unstable since the secular terms grow
without bound as $t\rightarrow \infty $. Let us note that we have already
proven that any solution that does not coincide with the vortex will be
unstable (see the text between Eqs. (\ref{CDM90}) and (\ref{CDM100}) ). On
top of that, we have just now proven the linear instability of the power-law
model. However, this model satisfies (\ref{CDM110}) under $0.5<\varkappa <1$
and, hence, a Rayleigh stable flow may be unstable in the sense of Lyapunov
stability \cite{Michel 2008}. Let us now consider the solution, (\ref{LI10}%
)-(\ref{LI20}), to the system (\ref{NMD10})-(\ref{NMD20}) under $\varkappa =1
$. In such a case rank$\mathbf{A}=2$ since $A_{32}=0$ and, hence, the
solution to (\ref{LI180}) does not contain secular terms. Thus, this
solution will be unstable \cite{Michel 2008} if
\begin{equation}
A_{12}A_{21}>0.  \label{LI210}
\end{equation}
By virtue of (\ref{ASE25}) and (\ref{LI30}), we find that (\ref{LI210}) will
be valid if
\begin{equation}
r>\max \left( \frac{r_{m}}{1-E_{u}F_{r}},\frac{r_{m}}{\left(
2E_{u}F_{r}-1\right) }\right) .  \label{LI220}
\end{equation}
Since the disk is thin, i.e. $E_{u}\ll 1$, we obtain the following condition
of linear instability:
\begin{equation}
r>\frac{r_{m}}{1-E_{u}F_{r}}.  \label{LI230}
\end{equation}
Thus, the vortex will be linearly unstable, at least, on the disk's
periphery, provided $R_{e}\rightarrow \infty $. Let us remind that the
vortex is unconditionally Rayleigh unstable because the Rayleigh frequency $%
f_{R}^{2}=0$.

\subsection{Magnetic accretion disk}

We will use the following asymptotic expansion in the limit $z\diagup
r\rightarrow 0$. Then, in view of the symmetry, we may write:
\begin{equation*}
B_{\varphi }=B_{\varphi 1}z+\ldots ,\ B_{z}=B_{z0}+B_{z2}z^{2}+\ldots ,\
B_{r}=B_{r1}z+B_{\varphi 3}z^{3}+\ldots ,\
\end{equation*}
\begin{equation*}
\rho =\rho _{0}+\rho _{2}z^{2}+\ldots ,\ T=T_{0}+T_{2}z^{2}+\ldots ,\
v_{\varphi }=v_{\varphi 0}+v_{\varphi 2}z^{2}+\ldots ,\
\end{equation*}
\begin{equation*}
v_{r}=v_{r0}+v_{r2}z^{2}+\ldots ,\ v_{z}=v_{z1}z+v_{z3}z^{3}+\ldots ,
\end{equation*}
\begin{equation}
\Phi =-\frac{1}{\sqrt{r^{2}+z^{2}}}=\Phi _{0}+\Phi _{2}z^{2}+\ldots =-\frac{1%
}{r}+\frac{1}{2r^{3}}z^{2}+\ldots \ ,  \label{SSMAD05}
\end{equation}
where all coefficients depend on $r$, only. It is also assumed in this
section that $B_{z0}\neq 0$.

A steady-state magnetic accretion disk with $v_{r0}\neq 0$, $\rho _{0}\neq 0$%
, and with a negligible dynamic viscosity, i.e. $\mu =0$, will be our
initial concern. The magnetic field will be called as almost poloidal if
\begin{equation}
B_{\varphi 1}=0\Rightarrow B_{\varphi }=O\left( z^{3}\right) ,
\label{SSMAD13}
\end{equation}
and the magnetic field will be called as almost axial if
\begin{equation}
B_{\varphi 1}=0,\ B_{r1}=0\Rightarrow B_{\varphi }=O\left( z^{3}\right) ,\
B_{r}=O\left( z^{3}\right) .  \label{SSMAD15}
\end{equation}

Let us prove that the circular velocity at the midplane will be the vortex
velocity, i.e.
\begin{equation}
v_{\varphi 0}\equiv \left. v_{\varphi }\right| _{z=0}=\frac{C_{\varphi 0}}{r}%
,\quad C_{\varphi 0}=const,  \label{SSMAD10}
\end{equation}
if and only if the magnetic field will be almost poloidal. Actually, since $%
\mu =0$, the steady-state version of Eqs. (\ref{EM240}), (\ref{EM270}) is
the following.
\begin{equation}
\frac{1}{r}\frac{\partial \left( r\rho v_{r}\right) }{\partial r}+\frac{%
\partial \left( \rho v_{z}\right) }{\partial z}=0,  \label{SSMAD20}
\end{equation}
\begin{equation*}
\frac{1}{r}\frac{\partial }{\partial r}r\left( \rho v_{\varphi }v_{r}-\frac{%
E_{u}}{\beta }B_{\varphi }B_{r}\right) +
\end{equation*}
\begin{equation}
\frac{\partial }{\partial z}\left( \rho v_{\varphi }v_{z}-\frac{E_{u}}{\beta
}B_{\varphi }B_{z}\right) +\frac{\rho v_{\varphi }v_{r}}{r}-\frac{E_{u}}{%
\beta }\frac{B_{\varphi }B_{r}}{r}=0.  \label{SSMAD40}
\end{equation}
In view of (\ref{SSMAD20}), (\ref{SSMAD40}), and (\ref{SSMAD05}), we obtain
\begin{equation}
\frac{1}{r}\frac{\partial \left( r\rho _{0}v_{r0}\right) }{\partial r}+\rho
_{0}v_{z1}=0,  \label{SSMAD100}
\end{equation}
\begin{equation}
\frac{v_{\varphi 0}}{r}\frac{\partial }{\partial r}r\left( \rho
_{0}v_{r0}\right) +\frac{r\left( \rho _{0}v_{r0}\right) }{r}\frac{\partial
v_{\varphi 0}}{\partial r}-\frac{E_{u}}{\beta }B_{\varphi 1}B_{z0}+\rho
_{0}v_{\varphi 0}v_{z1}+\frac{\rho _{0}v_{\varphi 0}v_{r0}}{r}=0.
\label{SSMAD110}
\end{equation}
By virtue of (\ref{SSMAD100}), we obtain from (\ref{SSMAD110}):
\begin{equation}
\rho _{0}v_{r0}\frac{\partial v_{\varphi 0}}{\partial r}-\frac{E_{u}}{\beta }%
B_{\varphi 1}B_{z0}+\frac{\rho _{0}v_{\varphi 0}v_{r0}}{r}=0.
\label{SSMAD120}
\end{equation}
Let the magnetic field will be almost poloidal, i.e. $B_{\varphi 1}=0$.
Then, in view of (\ref{SSMAD120}), we have
\begin{equation}
\rho _{0}v_{r0}\frac{\partial rv_{\varphi 0}}{r\partial r}=0.
\label{SSMAD130}
\end{equation}
Equality (\ref{SSMAD130}) proves (\ref{SSMAD10}).

Let (\ref{SSMAD10}) be valid. In view of (\ref{SSMAD120}), we have
\begin{equation}
\rho _{0}v_{r0}\frac{\partial rv_{\varphi 0}}{r\partial r}-\frac{E_{u}}{%
\beta }B_{\varphi 1}B_{z0}=0.  \label{SSMAD140}
\end{equation}
By virtue of (\ref{SSMAD10}), we find from (\ref{SSMAD140}):
\begin{equation}
B_{\varphi 1}B_{z0}=0.  \label{SSMAD150}
\end{equation}
Equality (\ref{SSMAD150}) proves (\ref{SSMAD13}).

Let us now consider a steady-state magnetic disk with $v_{r}=0$ (and, hence,
$v_{z}=0$) and with a negligible dynamic viscosity, i.e. $\mu =0$. Let us
remind that such assumptions for the case of non-magnetic disk lead to the
ill-posed problem in the sense of Hadamard (see Sec. \ref{Non-magnetic disk}%
). In particular, the solution to the mathematical model (Sec. \ref
{Non-magnetic disk}) is not unique. We intend to find a steady-state
solution to the MHD system (\ref{EM240})-(\ref{EM289}) provided $%
R_{e}\rightarrow \infty $, $\vartheta _{ke}=0$. In such a case, the MHD
system is reduced to the following.
\begin{equation*}
-\frac{1}{r}\frac{\partial }{\partial r}r\left( \frac{E_{u}}{\beta }%
B_{r}^{2}\right) -\frac{\partial }{\partial z}\left( \frac{E_{u}}{\beta }%
B_{r}B_{z}\right) +
\end{equation*}
\begin{equation}
\frac{E_{u}}{\beta }\frac{B_{\varphi }^{2}}{r}-\frac{\rho v_{\varphi }^{2}}{r%
}=-\frac{\partial }{\partial r}\left( E_{u}P+\frac{E_{u}}{\beta }\frac{B^{2}%
}{2}\right) -\frac{\rho }{F_{r}}\frac{\partial \Phi }{\partial r},
\label{SV10}
\end{equation}
\begin{equation}
-\frac{1}{r}\frac{\partial }{\partial r}r\left( \frac{E_{u}}{\beta }%
B_{\varphi }B_{r}\right) -\frac{\partial }{\partial z}\left( \frac{E_{u}}{%
\beta }B_{\varphi }B_{z}\right) -\frac{E_{u}}{\beta }\frac{B_{\varphi }B_{r}%
}{r}=0,  \label{SV20}
\end{equation}
\begin{equation*}
\frac{1}{r}\frac{\partial }{\partial r}r\left( \frac{E_{u}}{\beta }%
B_{z}B_{r}\right) +\frac{\partial }{\partial z}\left( \frac{E_{u}}{\beta }%
B_{z}^{2}\right) =
\end{equation*}
\begin{equation}
\frac{\partial }{\partial z}\left( E_{u}P+\frac{E_{u}}{\beta }\frac{B^{2}}{2}%
\right) +\frac{\rho }{F_{r}}\frac{\partial \Phi }{\partial z}  \label{SV30}
\end{equation}
\begin{equation}
\frac{\partial \left( v_{\varphi }B_{r}\right) }{\partial r}+\frac{\partial
\left( v_{\varphi }B_{z}\right) }{\partial z}=0,  \label{SV40}
\end{equation}
\begin{equation}
\frac{1}{r}\frac{\partial rB_{r}}{\partial r}+\frac{\partial B_{z}}{\partial
z}=0.  \label{SV50}
\end{equation}
Eq. (\ref{EMD320}) is reduced to the following
\begin{equation}
\frac{\partial r\left( v_{\varphi }B_{z}\right) }{r\partial r}-\frac{%
\partial \left( v_{\varphi }B_{r}\right) }{\partial z}=0.  \label{SV60}
\end{equation}
Let $B_{\varphi }$\ be an odd function of $z$ and let $B_{r}=O\left(
z^{5}\right) $. Then we find from (\ref{SV50}) that
\begin{equation}
B_{z0}=B_{z0}\left( r\right) ,\ B_{z2}=B_{z4}=0.  \label{SV70}
\end{equation}
By virtue of (\ref{SV70}), we find from (\ref{SV40}) and (\ref{SV20}) that
\begin{equation}
v_{\varphi 0}=v_{\varphi 0}\left( r\right) ,\ v_{\varphi 2}=v_{\varphi
4}=0,\ B_{\varphi 1}=B_{\varphi 3}=0.  \label{SV80}
\end{equation}
Taking into account that $B_{r}=O\left( z^{5}\right) $ and using (\ref{SV70}%
), we obtain from Eq. (\ref{SV30}) that (\ref{NM30}) is valid for the case
of the magnetic disk as well. Hence, using (\ref{NM70}), (\ref{NM80}), and (%
\ref{SV70})-(\ref{SV80}), we obtain from Eq. (\ref{SV10}) that
\begin{equation}
\frac{\rho v_{\varphi 0}^{2}}{r}=E_{u}T\frac{\partial \rho }{\partial r}%
+E_{u}\rho \frac{\partial T}{\partial r}+\frac{E_{u}}{\beta }\frac{\partial
B_{z0}^{2}}{\partial r}+\frac{\rho }{F_{r}}\frac{\partial \Phi }{\partial r}%
+O\left( z^{4}\right) .  \label{SV90}
\end{equation}
Eq. (\ref{SV90}) must be valid under all values of $z$. Assuming $z=H$, we
obtain, in view of (\ref{NM20}), the following equation in $B_{z0}$:
\begin{equation}
\frac{E_{u}}{\beta }\frac{\partial B_{z0}^{2}}{\partial r}+O\left( \left(
z\diagup r\right) ^{4}\right) =0.  \label{SV100}
\end{equation}
Then we find:
\begin{equation}
B_{z0}\approx const.  \label{SV110}
\end{equation}
Since $B_{r}=O\left( z^{5}\right) $, we obtain from (\ref{SV60}), by virtue
of (\ref{SV70})-(\ref{SV80}), that
\begin{equation}
\frac{\partial r\left( v_{\varphi 0}B_{z0}\right) }{r\partial r}=0.
\label{SV120}
\end{equation}
We find, in view of (\ref{SV110}), (\ref{SV120}), that
\begin{equation}
v_{\varphi 0}\approx \frac{const}{r}.  \label{SV130}
\end{equation}

Analogously, if $B_{r}=0$ (and $v_{r}=0$), then we find from (\ref{SV10})-(%
\ref{SV50}) that $B_{\varphi }=0$ and $B_{z}=const$. Thus, in view of (\ref
{SV60}), the vortex velocity distribution
\begin{equation}
v_{\varphi }=\frac{const}{r}  \label{SV140}
\end{equation}
will be the only solution for the circular velocity, and, hence, the motion
will be Rayleigh unstable. Let us note, the density and temperature
distributions can be easily found from Eqs. (\ref{SV10}), (\ref{SV30}),
which can be written, in view of the foregoing, as the following:
\begin{equation}
\frac{\rho _{0}v_{\varphi }^{2}}{r}=\frac{\partial }{\partial r}\left(
E_{u}\rho _{0}T_{0}\right) +\frac{\rho _{0}}{r^{2}F_{r}},  \label{SV150}
\end{equation}
\begin{equation}
\frac{\partial }{\partial z}\left( E_{u}\rho T\right) +\frac{\rho }{F_{r}}%
\frac{z}{\left( r^{2}+z^{2}\right) ^{3\diagup 2}}=0,  \label{SV160}
\end{equation}
where the circular velocity, $v_{\varphi }$, is calculated from Eq. (\ref
{SV140}). Notice, Eqs. (\ref{SV150}), (\ref{SV160}) coincide with Eqs. (\ref
{NMD10}), (\ref{NMD20}), and, hence, we obtain (\ref{NM70}), (\ref{NM80}),
and (\ref{NM120}).

Notice, we have assumed $v_{r0}\neq 0$ in the above-proven assertion that
the midplane circular velocity will be (\ref{SSMAD10}) if and only if the
magnetic field will be almost poloidal. The following counter-example
demonstrates that the condition $v_{r0}\neq 0$ is essential. We consider the
case when $v_{r}=0$ and $B_{z}=B_{z0}+O\left( z^{4}\right) $ with $%
B_{z0}=const$. In view of (\ref{SV50}), we have
\begin{equation}
B_{r1}=\frac{C_{r1}}{r},\quad C_{r1}=const.  \label{SV170}
\end{equation}
Then, by virtue of (\ref{SV170}), we obtain from (\ref{SV60}) the power-law
model:
\begin{equation}
v_{\varphi 0}=\frac{const}{r^{\varkappa }},\quad \varkappa \equiv 1-\frac{%
C_{r1}}{B_{z0}}=const,  \label{SV180}
\end{equation}
which is the only solution for the midplane circular velocity in the frame
of our assumptions. However, in view of (\ref{SV20}), $B_{\varphi 1}=0$,
i.e. the magnetic field is almost poloidal.

\subsubsection{Viscous disk\label{Viscous disk}}

Let us, first, demonstrate that the midplane circular velocity will be the
vortex one, (\ref{SSMAD10}), if the dynamic viscosity $\mu =const\neq 0$,
provided that the magnetic field will be almost axial. Actually, if $%
B_{\varphi }=O\left( z^{3}\right) $ and $B_{r}=O\left( z^{3}\right) $, then,
in view of (\ref{SSMAD05}), we obtain from the steady-state version of Eq. (%
\ref{EM270}) that
\begin{equation}
\rho _{0}v_{r0}\frac{\partial rv_{\varphi 0}}{r\partial r}=\frac{1}{R_{e}}%
\frac{\partial }{\partial r}\left[ \mu r\frac{\partial }{\partial r}\left(
\frac{v_{\varphi 0}}{r}\right) \right] +\frac{2\mu v_{\varphi 2}}{R_{e}}+%
\frac{2\mu }{R_{e}}\frac{\partial }{\partial r}\left( \frac{v_{\varphi 0}}{r}%
\right) .  \label{SSMAD170}
\end{equation}
In view of (\ref{EM289}) and (\ref{SSMAD05}), we have
\begin{equation}
\frac{1}{r}\frac{\partial rB_{r1}}{\partial r}+2B_{z2}=0.  \label{SSMAD175}
\end{equation}
Hence
\begin{equation}
B_{z2}=0.  \label{SSMAD180}
\end{equation}
By virtue of (\ref{SSMAD05}), we find from (\ref{EM288}) that
\begin{equation}
\frac{\partial \left( v_{r0}B_{\varphi 1}-v_{\varphi 0}B_{r1}\right) }{%
\partial r}+2\left( v_{z1}B_{\varphi 1}-v_{\varphi 2}B_{z0}-v_{\varphi
0}B_{z2}\right) =0.  \label{SSMAD185}
\end{equation}
Hence, in view of (\ref{SSMAD180}), we obtain:
\begin{equation}
v_{\varphi 2}B_{z0}=0\Rightarrow v_{\varphi 2}=0.  \label{SSMAD190}
\end{equation}
By virtue of (\ref{SSMAD170}) and (\ref{SSMAD190}), we find that
\begin{equation}
\rho _{0}v_{r0}\frac{\partial rv_{\varphi 0}}{r\partial r}=\frac{\mu }{R_{e}}%
\frac{\partial }{\partial r}\left[ r\frac{\partial }{\partial r}\left( \frac{%
v_{\varphi 0}}{r}\right) +2\frac{v_{\varphi 0}}{r}\right] .  \label{SSMAD200}
\end{equation}
Obviously, the vortex velocity, (\ref{SSMAD10}), fulfills Eq. (\ref{SSMAD200}%
), no matter whether $v_{r0}\neq 0$ or $v_{r0}=0$. Hence, in contrast to the
non-viscous disk, we obtain the only solution, (\ref{SSMAD10}), to Eq. (\ref
{SSMAD200}) provided that $v_{r0}=0$.

As indicated above, the midplane circular velocity can be the vortex
velocity if the dynamic viscosity $\mu =const$. Let us now consider the case
when the midplane circular velocity is not the vortex velocity, but the
power-law model:
\begin{equation}
v_{\varphi 0}=\frac{C_{\varphi }}{r^{\varkappa }},\quad \varkappa =const\neq
1,\ C_{\varphi }=const.  \label{SSMAD210}
\end{equation}
The power-law model, (\ref{SSMAD210}), can fulfill Eq. (\ref{SSMAD200}) if $%
\mu =\mu \left( r\right) $. The following inequality must be valid to fulfil
(\ref{NM140}).
\begin{equation}
\varkappa >0.5.  \label{SSMAD215}
\end{equation}
It is also assumed that the magnetic field will be almost axial, i.e. $%
B_{\varphi }=O\left( z^{3}\right) $ and $B_{r}=O\left( z^{3}\right) $. To
estimate $\mu =\mu \left( r\right) $ at $z=0$ we assume $v_{r0}=0$. We
obtain from the steady-state version of Eq. (\ref{EM270}) that
\begin{equation}
\frac{1}{R_{e}}\frac{\partial }{\partial r}\left[ \mu r\frac{\partial }{%
\partial r}\left( \frac{v_{\varphi 0}}{r}\right) \right] +\frac{2\mu }{R_{e}}%
\frac{\partial }{\partial r}\left( \frac{v_{\varphi 0}}{r}\right) =0.
\label{SSMAD220}
\end{equation}
Then, by virtue of (\ref{SSMAD210}), we find from (\ref{SSMAD220}) that
\begin{equation}
\mu =\frac{C_{\mu }}{r^{1-\varkappa }},\quad C_{\mu }=const.
\label{SSMAD230}
\end{equation}
Assuming the characteristic quantity, $\mu _{\ast }$, for viscosity such
that $C_{\mu }=1$ and using (\ref{ASE25}), we represent the viscosity, $\mu $%
,\ in the following form:
\begin{equation}
\mu =T^{1-\varkappa }.  \label{SSMAD240}
\end{equation}

The power law (e.g., \cite{Carlson 1998}, \cite{Loitsyanskiy 1978}, \cite
{White 2006}) for the laminar viscosity, $\mu _{l}$, of dilute gases can be
written as the following
\begin{equation}
\mu _{l}=T^{\theta },  \label{SSMAD250}
\end{equation}
where typically $\theta =0.76$ \cite{Carlson 1998}. It is, also, assumed
that $\theta =8\diagup 9$ if $90<T<300$ $^{\circ }K$, and $\theta =0.75$ if $%
250<T<600$ $^{\circ }K$. If we assume that the flow in question is laminar,
then $\mu =\mu _{l}$, and, hence, $\varkappa =1-\theta $. Since we consider
the flow at the periphery of the disk, i.e. under a low temperature, we find
that
\begin{equation}
\varkappa <0.25.  \label{SSMAD260}
\end{equation}
The inequality (\ref{SSMAD260}) contradicts to (\ref{SSMAD215}), and, hence,
in contrast to the non-viscous disk under $v_{r0}=0$, the power-law model, (%
\ref{SSMAD210}), can not be assumed as a correct midplane circular velocity
in the case of laminar viscous flow, provided that $T\rightarrow 0$ as $%
r\rightarrow \infty $ (namely, $T\propto r^{-1}$).

Let us now consider a possibility for the power-law model, (\ref{SSMAD210}),
to be a correct midplane circular velocity provided the flow is turbulent.

The Euler number characterizes ``losses'' in a flow \cite{Horneber et al.
2012}, and it is higher in a turbulent flow than that in the laminar regime
\cite{Bhoite and Narasimham 2009}. Let us now estimate the value of $%
\varkappa $ in (\ref{SSMAD210}) by assuming that the turbulent gas tends to
flow with minimal losses, i.e. to have the Euler number as small as
possible. Using Prandtl and Kolmogorov suggestion \cite[p. 230]{Anderson et
al. 1984} that the turbulent viscosity, $\mu $, is proportional to the
square root of the kinetic energy of turbulence, $\overline{\kappa }$, we
evaluate $\mu _{0}\equiv \left. \mu \right| _{z=0}$ as
\begin{equation}
\mu _{0}=C_{\kappa }L_{\kappa }\rho _{0}\overline{\kappa }^{0.5},\quad
C_{\kappa },L_{\kappa }=const.  \label{SSMAD270}
\end{equation}
Let the disk be non-magnetic, $v_{r0}=0$, the kinetic energy of turbulence $%
\overline{\kappa }=const$, and let $\overline{\kappa }_{\ast }\ll RT_{\ast }$
(i.e., $\vartheta _{ke}\ll E_{u}$). Then, by virtue of (\ref{ASE25}), (\ref
{NM185}), and (\ref{SSMAD210}), we rewrite (\ref{NM120}) to read:
\begin{equation}
\rho _{0}=\frac{\rho _{0}^{0}}{r^{\alpha -1}}\exp \zeta \left[ \frac{1}{%
r^{2\varkappa -1}}-1\right] ,\quad \zeta =\frac{C_{\varphi }^{2}}{%
E_{u}\left( 2\varkappa -1\right) },\ \alpha =\frac{1}{E_{u}F_{r}}.
\label{SSMAD280}
\end{equation}
We obtain from (\ref{SSMAD270}) that
\begin{equation}
\mu _{0}=\frac{C_{\rho }}{r^{\alpha -1}}\exp \zeta \left[ \frac{1}{%
r^{2\varkappa -1}}-1\right] ,\quad C_{\rho }\equiv \rho _{0}^{0}C_{\kappa
}L_{\kappa }\overline{\kappa }^{0.5}=const.  \label{SSMAD290}
\end{equation}
Equating (\ref{SSMAD230}) and (\ref{SSMAD290}) at $r=1$, we find that $%
C_{\rho }=C_{\mu }$. Let us now assume that $\mu $ of (\ref{SSMAD230}) and $%
\mu _{0}$ of (\ref{SSMAD290}) coincide each other in the vicinity of $r=1$,
i.e. at $r=1+\varepsilon $ ($\varepsilon \ll 1$), with accuracy $O\left(
\varepsilon ^{2}\right) $. In such a case we obtain that
\begin{equation}
E_{u}\propto \frac{1}{2-\varkappa }.  \label{SSMAD300}
\end{equation}
As we can see from (\ref{SSMAD300}) and (\ref{SSMAD215}), $E_{u}\rightarrow
\min $ if $\varkappa \rightarrow 0.5$. Thus, in the frame of our
assumptions, we find that the turbulent flow tends to be Keplerian.

\section{Concluding remarks}

On the basis of the exact solution to the MHD system in Sec. \ref{VID} we
may conclude that vertically isothermal disks will, in fact, be totally
isothermal under the assumption that the radial components, $v_{r}$ and $%
B_{r}$, of velocity and magnetic field, respectively, are negligible.
Furthermore, the disks cannot be considered as thin in terms of, e.g., \cite
{Duric 2004} even if the Euler number $0<E_{u}\ll 1$.

The exact solution to the MHD system in Sec. \ref{Adiabatic flow}
corroborates the view \cite{Shapiro and Teukolsky 2004} that thin disk
accretion must be highly nonadiabatic. Despite of the fact that adiabatic
disks (see Sec. \ref{Adiabatic flow}) are more trustworthy than isothermal
ones, we find that the non-dimensional semi-thickness $H\propto r^{2}$
instead of the ratio $H\propto r$ \cite{Shapiro and Teukolsky 2004}.

The exact solutions to the MHD systems in Sec. \ref{VID} and Sec. \ref
{Adiabatic flow} prove that the vortex velocity will be the only solution
for the circular velocity provided that the flow is charge-neutral. Let us
note that the exact solutions are found under the assumption that $v_{r}=0$
and $B_{r}=0$.

The approach developed in Sec. \ref{Perfect gas} for the modeling of thin
accretion disks turns out to be efficient. In the case of non-magnetic disk,
this approach enables to obtain, with ease: the solution for the
steady-state non-viscous disk with a good accuracy, to find the
non-dimensional semi-thickness $H\propto r$, to prove that all solutions for
the midplane circular velocity are unstable provided the disk is
non-viscous. Using this approach one can prove (under $v_{r}\neq 0$) that
the midplane circular velocity will be the vortex velocity if and only if
the magnetic field will be almost poloidal. The approach of Sec. \ref
{Perfect gas} enables one to demonstrate that the pure hydrodynamic
turbulence\ in accretion disks is possible, and to demonstrate that the
turbulent flow tends to be Keplerian.

\end{document}